\documentclass[sigplan,10pt]{acmart}

\makeatletter

\def\@mkauthors{%
  \begingroup
  \global\setbox\mktitle@bx=\vbox{%
    \unvbox\mktitle@bx%
    \par\medskip
    \hsize=\textwidth
    \centering\normalfont\normalsize
    {\large
      Shaoyu Wang\textsuperscript{1}\quad
      Yizhuo Liang\textsuperscript{1}\quad
      Jaeyong Song\textsuperscript{2}\quad
      Chong Li\textsuperscript{3}\quad
      Seo Jin Park\textsuperscript{1}\par}%
    \vskip 5pt
    \textsuperscript{1}University of Southern California\quad
    \textsuperscript{2}Seoul National University\quad
    \textsuperscript{3}Point72\par
    \vskip 3pt
    {\par}%
    \bigskip}%
  \endgroup}
\makeatother

\renewcommand\footnotetextcopyrightpermission[1]{}

\AtBeginDocument{%
  }

\usepackage{amsfonts}
\usepackage[T1]{fontenc}
\usepackage[utf8]{inputenc}

\usepackage{upquote}

\usepackage{hyperref}
\usepackage{graphicx,grffile}

\usepackage{multirow}
\usepackage{xspace}
\usepackage{tabularx}
\usepackage{ragged2e}
\usepackage{booktabs}
\usepackage{paralist}
\usepackage[american]{babel}
\usepackage{courier,color,wrapfig}
\usepackage{balance}
\usepackage{enumitem}
\setitemize{topsep=3pt,partopsep=0pt,noitemsep,parsep=0pt,leftmargin=15pt}
\usepackage{epstopdf}
\usepackage{multirow}
\usepackage{booktabs}
\usepackage{url}
\usepackage{pifont}
\usepackage{subfigure}
\usepackage{tikz}
\usepackage{comment}
\usepackage{mathtools}
\usepackage[normalem]{ulem}
\usepackage{xkeyval}
\usepackage{threeparttable}
\usepackage{makecell}

\usepackage{algorithm}
\usepackage{algpseudocode}
\usepackage{etoolbox}
\usepackage{ifthen}
\usepackage{sepfootnotes}

\usepackage[capitalize,noabbrev]{cleveref}
\usepackage{wrapfig}
\usepackage{textcomp}
\usepackage{tcolorbox}
\usepackage{newunicodechar}

\usepackage{listings}
\usepackage{xcolor}

\usepackage{titlesec}
\titlespacing*{\section}{0pt}{7pt plus 3pt minus 3pt}{3pt plus 3pt minus 2pt}
\titlespacing*{\subsection}{0pt}{4pt plus 3pt minus 2pt}{1pt plus 3pt minus 1pt}
\titlespacing*{\subsubsection}{0pt}{4pt plus 3pt minus 2pt}{0pt plus 3pt minus 1pt}
\titleformat{\section}{\large\bfseries}{\thesection}{1em}{}
\titleformat{\subsection}{\normalsize\bfseries}{\thesubsection}{1em}{}

\makeatletter
\let\ACM@origsection\section
\let\ACM@origsubsection\subsection
\let\ACM@origsubsubsection\subsubsection
\let\ACM@origparagraph\paragraph
\makeatother

\newcommand\paraspace{\vspace*{0.5ex}}
\providecommand\parab[1]{\paraspace\noindent\textbf{#1}}

\newcommand{\sysname}{Moebius\xspace}

\newcommand{\secref}[1]{\S\ref{#1}}
\newcommand{\appref}[1]{Appendix~\ref{#1}}
\newcommand{\figref}[1]{Figure~\ref{#1}}
\newcommand{\tabref}[1]{Table~\ref{#1}}

\definecolor{teal}{rgb}{0.0, 0.5, 0.5}
\definecolor{olive}{rgb}{0.5, 0.5, 0.0}
\definecolor{pink}{rgb}{1.0, 0.75, 0.8}
\definecolor{lightgray}{gray}{0.75}
\definecolor{mediumgray}{gray}{0.5}
\definecolor{darkgray}{gray}{0.25}
\definecolor{charcoal}{gray}{0.2}
\definecolor{turquoise}{rgb}{0.25, 0.88, 0.82}
\definecolor{coral}{rgb}{1.0, 0.5, 0.31}
\definecolor{navyblue}{rgb}{0.0, 0.0, 0.5}
\definecolor{lime}{rgb}{0.75, 1.0, 0.0}
\definecolor{darkgreen}{rgb}{0.0, 0.5, 0.0}
\definecolor{violet}{rgb}{0.56, 0.0, 1.0}
\definecolor{lightgreen}{rgb}{0.85, 1.0, 0.85}
\definecolor{blue(ncs)}{rgb}{0.0, 0.53, 0.74}

\definecolor{burgundy}{cmyk}{0.5, 1.0, 0.7, 0.4}
\definecolor{olivegreen}{cmyk}{0.64, 0, 0.95, 0.4}
\definecolor{peach}{cmyk}{0, 0.5, 0.7, 0}
\definecolor{mustard}{cmyk}{0, 0.3, 1, 0}

\newtoggle{showmarks}
\toggletrue{showmarks}
\iftoggle{showmarks}{
\newcommand\red[1]{\textcolor{red}{#1}}
\newcommand\redstrike[1]{\red{\sout{#1}}}
\newcommand\green[1]{\textcolor{\green}{#1}}
\newcommand\greenstrike[1]{\green{\sout{#1}}}
\newcommand\orange[1]{\textcolor{orange}{#1}}
\newcommand\orangestrike[1]{\orange{\sout{#1}}}
\newcommand\blue[1]{\textcolor{blue}{#1}}
\newcommand\bluencs[1]{\textcolor{blue(ncs)}{#1}}
\newcommand\bluestrike[1]{\blue{\sout{#1}}}
\newcommand\purple[1]{\textcolor{purple}{#1}}
\newcommand\purplestrike[1]{\purple{\sout{#1}}}
\newcommand\teal[1]{\textcolor{teal}{#1}}
\newcommand\tealstrike[1]{\teal{\sout{#1}}}
\newcommand\turquoise[1]{\textcolor{turquoise}{#1}}
\newcommand\turquoisestrike[1]{\turquoise{\sout{#1}}}
\newcommand\darkgreen[1]{\textcolor{darkgreen}{#1}}
\newcommand\darkgreenstrike[1]{\darkgreen{\sout{#1}}}
\newcommand\lime[1]{\textcolor{lime}{#1}}
\newcommand\limestrike[1]{\lime{\sout{#1}}}
\newcommand\olivegreen[1]{\textcolor{olivegreen}{#1}}
\newcommand\olivegreenstrike[1]{\olivegreen{\sout{#1}}}

\newcommand{\seojin}[1]{[\orange{\sf\textit{#1 - Seojin}}]}

\newcommand{\JY}[1]{[\bluencs{\sf\textit{#1 - Jaeyong}}]}

\newcommand{\draft}[1]{\turquoise{\sf\textit{#1}}}
\newcommand{\todo}[1]{[\red{\sf\textbf{TODO: }\textit{#1}}]}
}{
\newcommand\red[1]{#1}
\newcommand\redstrike[1]{\unskip}
\newcommand\green[1]{#1}
\newcommand\greenstrike[1]{\unskip}
\newcommand\orange[1]{#1}
\newcommand\orangestrike[1]{\unskip}
\newcommand\blue[1]{#1}
\newcommand\bluestrike[1]{\unskip}
\newcommand\purple[1]{\unskip}
\newcommand\purplestrike[1]{\unskip}
\newcommand\teal[1]{\unskip}
\newcommand\tealstrike[1]{\unskip}
\newcommand\turquoise[1]{\unskip}
\newcommand\turquoisestrike[1]{\unskip}
\newcommand\darkgreen[1]{\unskip}
\newcommand\darkgreenstrike[1]{\unskip}
\newcommand\lime[1]{\unskip}
\newcommand\limestrike[1]{\unskip}
\newcommand\olivegreen[1]{\unskip}
\newcommand\olivegreenstrike[1]{\unskip}

\newcommand{\seojin}[1]{}
\newcommand{\yibo}[1]{}
\newcommand{\junzhou}[1]{}
\newcommand{\JY}[1]{}

\newcommand{\draft}[1]{}
\newcommand{\todo}[1]{}
}

\begin{document}

\title{\textsc{\sysname}: Serving Mixture-of-Expert Models \\ with Seamless Runtime Parallelism Switch}





\begin{abstract}

Mixture-of-Experts (MoE) architectures scale large language models (LLMs) to hundreds of billions of parameters. Serving a single MoE model requires multiple GPUs operating in parallel, typically through tensor parallelism (TP) or expert parallelism (EP). The optimal choice depends on the number of in-flight requests: TP is faster at low concurrency, whereas EP wins at high concurrency. Production workloads cross this boundary continually: online serving sees bursty arrivals that subside into quiet periods, and reinforcement-learning rollouts begin as a high-concurrency burst that decays into a long tail of stragglers. Pinning either layout therefore forfeits performance when the workload crosses to the other side.

We present \sysname, a serving system that switches between EP and TP at runtime without restarting the engine or dropping in-flight requests.
Our key insight is that EP and TP are two layouts of one model, not two models: they compute the same function over byte-identical expert weights and KV cache, so a switch changes only which rank owns each slice.
Moving those owner-changed slices is the sole irreducible cost, and modern high-bandwidth GPU interconnects make it fast enough to do between decode steps without draining in-flight requests.
\sysname preserves each parallelism's runtime resident, and reshards the single copy of expert weights and KV cache at fixed addresses with fused GPU-to-GPU transfer kernels.
On 8$\times$H200 GPUs serving Qwen3-235B-A22B, \sysname matches the better static parallelism at every operating point, and beats it on RL rollouts by 1.16--1.25$\times$ across steps.
Each switch completes in 215--434\,ms, and \sysname holds both layouts resident with only 2.4\% memory overhead.
\end{abstract}



\settopmatter{printfolios=true, printacmref=false}
\pagestyle{plain}

\maketitle

\section{Introduction}
\label{sec:intro}

\begin{figure*}[t]
	\centering
	\begin{minipage}[t]{0.3\textwidth}
		\centering
		\includegraphics[width=\linewidth]{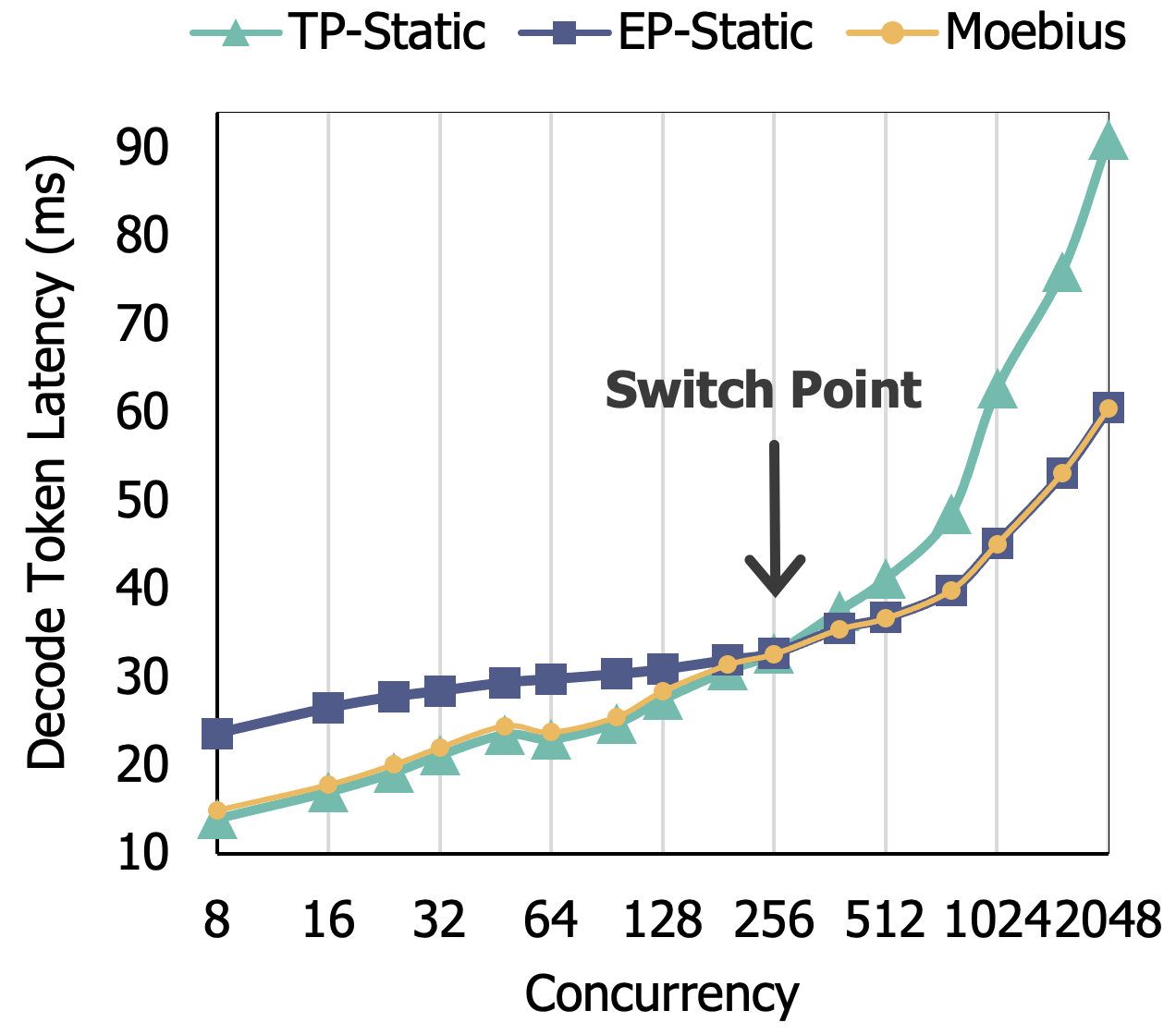}\\[2pt]
		{\small (a) Static load sweep}
		\label{fig:teaser-switch}
	\end{minipage}
	\hfill
	\begin{minipage}[t]{0.3\textwidth}
		\centering
		\includegraphics[width=\linewidth]{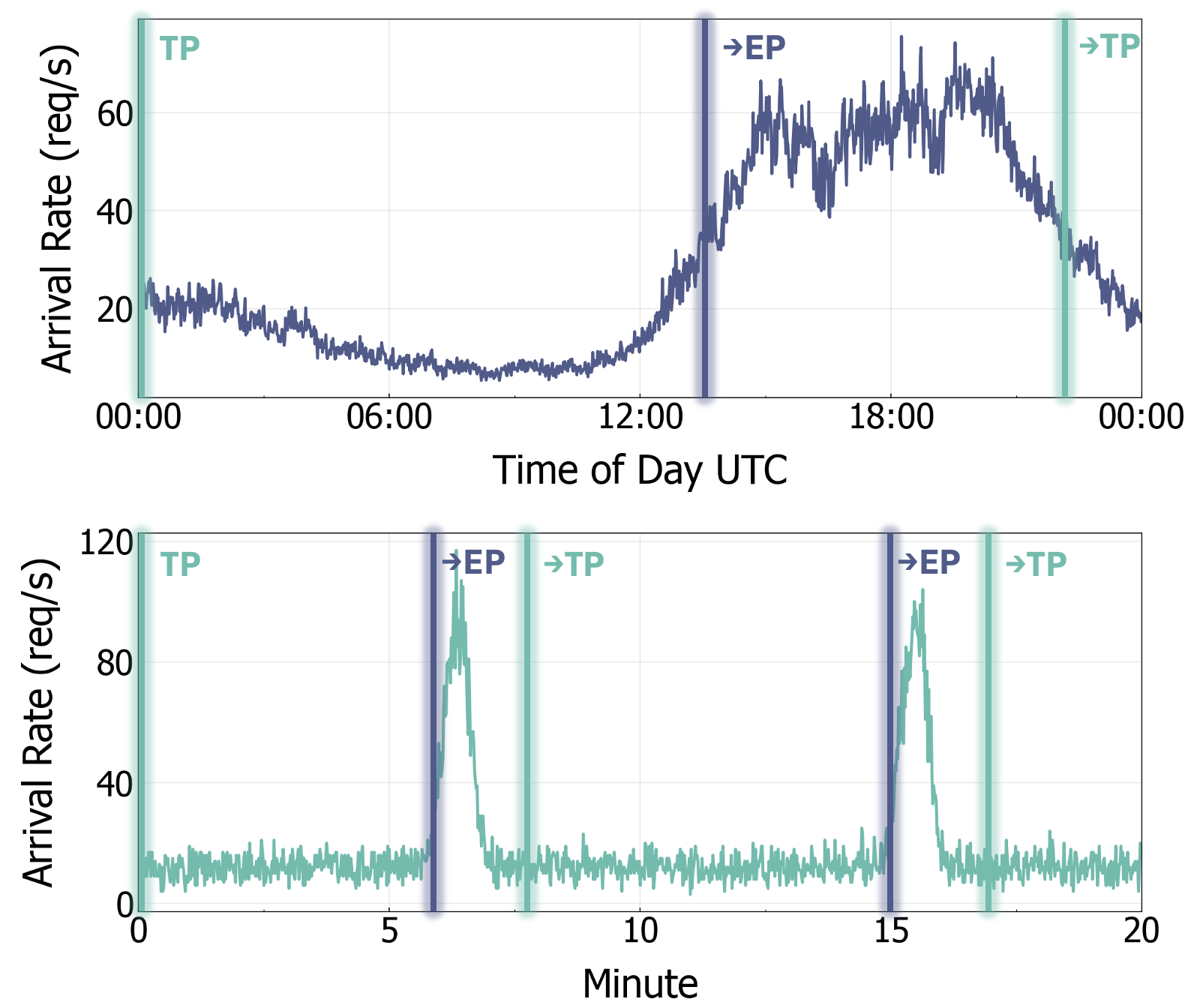}\\[2pt]
		{\small (b) Dynamic request loads}
		\label{fig:teaser_burst}
	\end{minipage}
	\hfill
	\begin{minipage}[t]{0.37\textwidth}
		\centering
		\includegraphics[width=\linewidth]{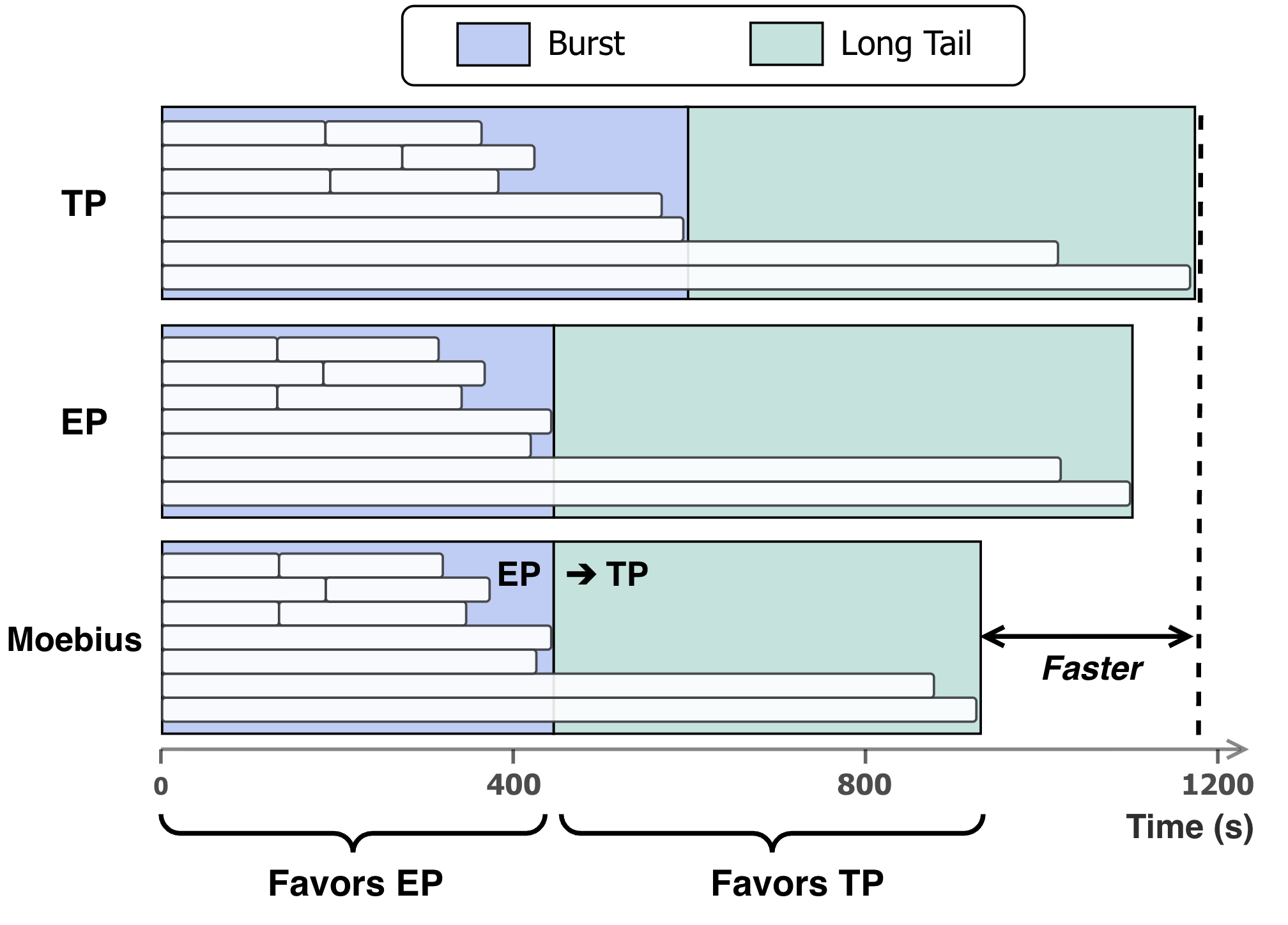}\\[2pt]
		{\small (c) RL rollout}
		\label{fig:teaser-rollout}
	\end{minipage}
	\caption{%
		Optimal parallelism for MoE decoding shifts with the active load.
		(a) Measured decode latency vs concurrency (i.e. global batch size) for TP, EP, and \sysname on a static load sweep (8\(\times\)H200, Qwen3-235B).
		The switch point marks the TP--EP crossover.
		(b) Request arrival rate (req/s) over time on an Azure online serving trace~\cite{azurellmtrace} (top) and a bursty trace (bottom).
		Vertical markers indicate switch points between TP and EP.
		(c) The same RL rollout batch runs on TP, EP, and \sysname.
		Each white strip is one sample's decode lifetime, with length proportional to the number of its output tokens.
		  An initial burst transitions to the long tail phase when most samples finish and only a few stragglers are still running.
		  Background shading marks the two phases, and \sysname switches EP$\rightarrow$TP at the boundary.
	}
	\label{fig:teaser}
\end{figure*}

Mixture-of-experts (MoE) \cite{shazeer2017outrageouslylargeneuralnetworks} has become a leading architecture for scaling large language models (LLMs).
By activating only a small subset of expert sub-networks per token, an MoE model grows to hundreds of billions of parameters without a proportional increase in per-token computation.
Recent frontier models such as Qwen-MoE~\cite{qwen3}, DeepSeek~\cite{deepseekv3}, and GPT-OSS~\cite{gptoss}, together with the serving systems~\cite{sglang, vllm, tensorrtllm} that efficiently host them, have made MoE inference a mainstream deployment target.

During MoE inference serving, a parallelism configuration defines how the MoE model weights are distributed across GPUs. \emph{Tensor parallelism (TP)} shards each expert and the attention heads across all GPUs, runs the full batch on every GPU, and synchronizes partial results with per-step \texttt{All-Reduce} communication.
\emph{Expert parallelism (EP)} assigns each GPU a subset of experts together with a disjoint slice of requests, routes tokens to their expert-owning GPUs with \texttt{All-to-All} dispatch, and combines the results. 


We observe that the optimal parallelism is critically dependent on the serving concurrency, as demonstrated in \figref{fig:teaser}(a).
A fixed parallelism is not uniformly preferable.
At low concurrency, EP fragments the batch across GPUs and leaves each rank with too little work to keep its kernels and attention efficient.
At high concurrency, TP suffers both from the per-layer \texttt{All-Reduce} on the full hidden state and from larger per-rank MoE activation traffic.
Real-world workloads vary in concurrency over time, requiring runtime switching.
Online serving traces exhibit bursty arrivals followed by quiet periods, repeatedly carrying the active batch across the TP--EP crossover and back (\figref{fig:teaser}(b)).
Beyond online serving, reinforcement learning (RL) rollouts such as GRPO~\cite{grpo} and DAPO~\cite{dapo} exhibit an even sharper variation: each rollout begins as a high-concurrency burst that favors EP and then decays into a long tail of stragglers that favors TP as sequences finish at different lengths~\cite{hu2024openrlhf, hybridflow} (\figref{fig:teaser}(c)).
Applying a fixed TP or EP layout to such dynamic workloads wastes the advantages of the other.



One might hope to reuse the runtime-switching approaches developed for dense LLM serving~\cite{shiftparallelism, amoeba, flyingserving}.
These systems, however, assume weights of both parallelisms are local on device, in-flight requests can be paused or drained, and runtime is cheap enough to rebuild. A production-level MoE serving system denies these assumptions, posing the following challenges for workload-adaptive parallelism switching.

\begin{itemize}[leftmargin=*]
    \item \textbf{Larger weight transfer.} Expert weights are much larger than those of dense LLMs. The weights of only one parallelism layout can stay alive on device. Thus, a switch must reshard GPU-resident experts across the interconnect. 
    \item  \textbf{Requests resume.} EP partitions requests across GPUs, while TP makes every GPU serve the full requests with sharded heads; therefore, a live switch must also redistribute in-flight requests and their paged key-value (KV) cache~\cite{vllm} to resume the decoding.
    \item \textbf{Graph-based execution.} Nowadays, graph-based execution with pre-capturing (i.e., CUDA graph)  is a de facto standard for low-latency serving.
    Tensor addresses are embedded in the graph, and a switch must not rewrite the captured addresses of the weights and cache.
\end{itemize}



To address these challenges, we present \sysname, the first MoE serving system that switches between EP and TP at runtime without draining in-flight requests.
At first glance, such a switch looks expensive on every count: it appears to demand replicating/reloading gigabyte-scale expert weights, redistributing live requests, and recapturing CUDA graphs on every layout change.

Our key insight is that on modern high-bandwidth GPU interconnects such as NVLink, expert weights need no longer be treated as fixed assets: EP and TP are two \emph{layouts of one model}, computing the same function over byte-identical weights and KV, and differing only in how that state is sharded across GPUs and which collectives move it.
A parallelism switch is therefore not intrinsically expensive: its only irreducible cost is moving the data whose owners change.
\sysname minimizes the cost: it keeps the control plane (CUDA graphs, communication groups, attention metadata) resident for both layouts, and reshards the data plane (expert weights, paged KV cache) over a single fixed-address allocation, so a switch \emph{selects} the prepared runtime and moves \emph{only} the owner-changed bytes.

\sysname realizes this design with three mechanisms.
To manage the unchanged shared state between EP and TP, we use a \textit{unified memory manager}.
It pre-allocates contiguous storage for expert weights, KV cache, request buffers, and transfer scratch, exposing stable parallelism-specific tensor views, so captured CUDA graphs remain valid across switches.
To provide a lightweight switch while supporting redistribution of in-flight requests, \textit{fused direct-transfer kernels} move expert slices and KV shards directly into peer-GPU destination slots over GPU network, avoiding staged collective buffers and auxiliary on-device memory copy.
Integrated with the above components, to support graph-based execution for low-latency serving, \textit{runtime-preservation} mechanisms keep both parallelisms' CUDA graphs, communication groups, and attention metadata resident, so that a switch selects prepared runtime state rather than rebuilding it.

We benchmark \sysname on 8$\times$H200 GPUs serving Qwen3-235B-A22B. 
Across a concurrency sweep of MoE serving as a microbenchmark, \sysname matches the better static layout at every operating point, as briefly shown in \figref{fig:teaser}(a).
On dynamic workloads, \sysname provides significant performance improvement with low switching overhead.
On RL rollouts, it is the fastest at every rollout step, beating even an oracle that picks the better static layout per step by 1.16--1.25$\times$, and either fixed baseline by up to 1.31$\times$.
On a bursty online-serving trace, it switches frequently between TP and EP with low overhead, improving both the prefill-bound time-to-first-token during bursts and the decode-bound time-per-output-token in the quiet periods. 
In these benchmarks, each switch completes in 215--434\,ms.
The direct-transfer kernel reshards expert weights in 152\,ms, 1.49$\times$ faster than an NCCL collective.
Holding both layouts resident stays below static TP and within 0.2\,GB of static EP, its 2.4\% dual-mode buffer funded by shrinking the KV cache rather than adding memory. 

This paper makes the following contributions:
\begin{itemize}[leftmargin=*]
	\item 
    We show that the optimal parallelism layout crosses between TP and EP \emph{within a single serving episode}, across a load sweep and through the burst-to-tail decay in RL rollouts. Parallelism must therefore be chosen at runtime rather than fixed at deployment, yet no existing serving system switches between EP and TP live during decode.
	\item 
    To support dynamic switching between EP and TP, we define a bidirectional transformation that reshards expert weights and migrates in-flight requests with their paged KV cache across the two layouts' different attention shardings, producing outputs equivalent to those the destination layout would produce without draining the engine or recomputing any request.
	\item 
    We design, implement, and evaluate \sysname, which realizes the transformation through fixed-address unified memory buffer that keeps captured CUDA graphs valid, direct GPU-to-GPU transfer kernels that reshard $1.49\times$ faster than an NCCL collective, and resident dual runtimes that are selected rather than rebuilt, switching in 215--434\,ms at roughly 2.4\% memory overhead.
\end{itemize}


\section{Background and Motivation}
\label{sec:background}

\subsection{Parallelism for MoE Inference}
\label{sec:bg-parallelism}

MoE inference factors into two independent placement choices.
Attention can be \textit{tensor-parallel} (TP), where every GPU holds the full batch with sharded heads, or \textit{data-parallel} (DP), where each GPU owns a disjoint request slice and stores the KV cache for only that slice.
Experts can be \textit{tensor-parallel}, where each weight matrix is sharded with \texttt{All-Reduce}~\cite{megatronTP}, or \textit{expert-parallel} (EP), where each GPU owns a subset of experts and tokens are routed via \texttt{All-to-All}~\cite{DeepseekAIDeepEP, ucclep, PerplexityAIpplx-kernels}.
The cross product gives four layouts, written attention/expert: TP/TP, DP/EP, DP/TP, and TP/EP.
TP attention also replicates the KV cache when the model has fewer KV heads than TP ranks, capping request capacity.

\begin{figure}[t]
    \centering
    \includegraphics[width=0.85\columnwidth]{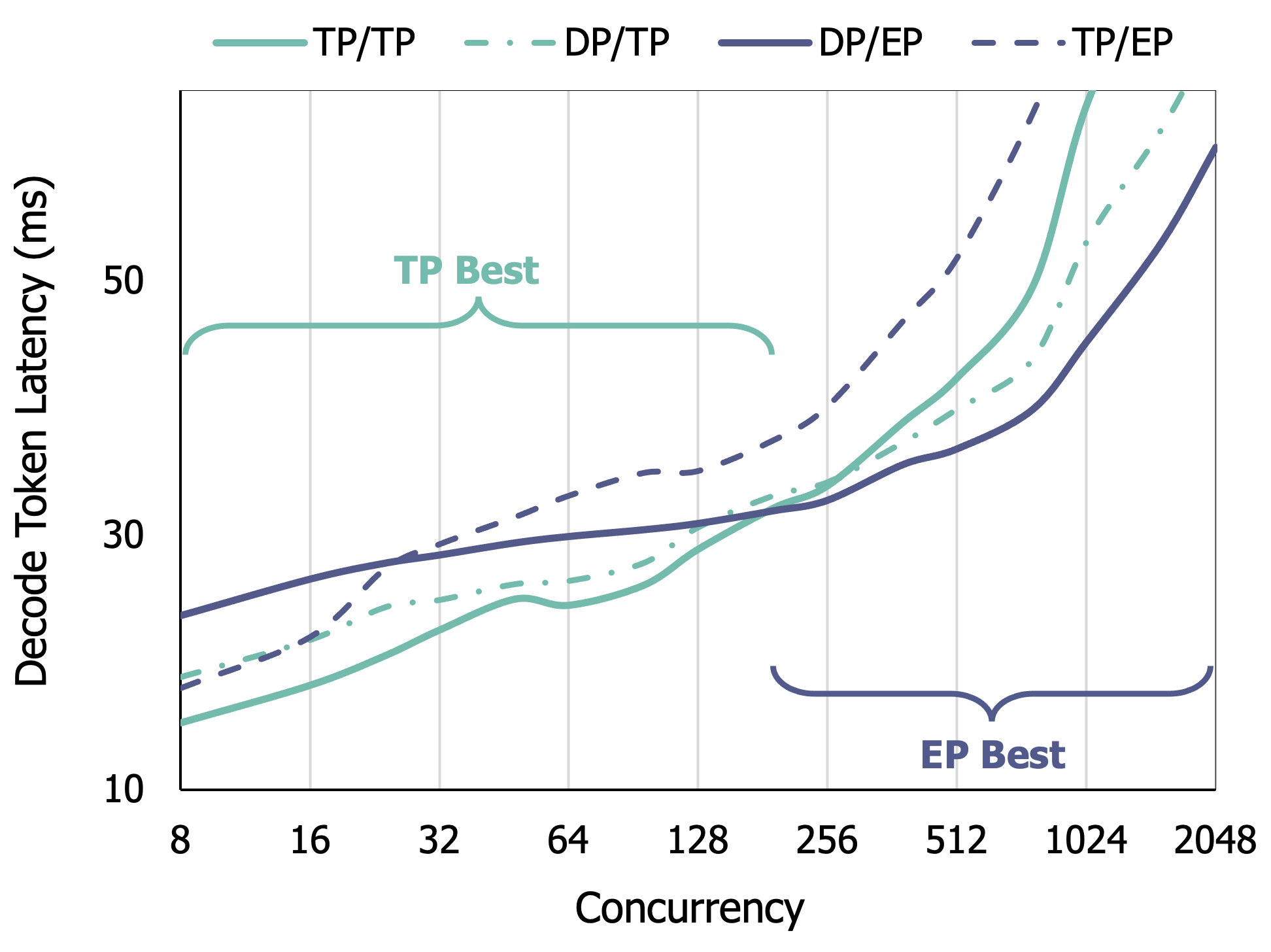}
    \caption{Steady-state decode-step latency for Qwen3-235B-A22B on 8\(\times\)H200 GPUs.
    Lower is better.
    }
    \label{fig:bg-latency}
\end{figure}

\parab{The favorable layout depends on load.}
We swept steady-state decode concurrency on Qwen3-235B-A22B across 8\(\times\)H200 GPUs with CUDA graphs enabled, measuring per-step decode latency for each of the four layouts.
Across global batch sizes \(B\) from 8 to 2048, TP/TP won by \(1.5\times\) at small batches and DP/EP won by \(1.5\times\) at large batches (\figref{fig:bg-latency}), with the boundary between \(B{=}128\) and \(B{=}256\).
The mixed layouts never joined the lower envelope, for structural reasons: DP/TP gathers the full token set before running TP experts and so misses DP/EP's MoE-volume reduction, while TP/EP keeps TP attention's full-token batch and feeds every token through the routing layer, erasing the activation-volume savings EP exists to provide.
The rest of the paper therefore focuses on the two survivors, which we write \textbf{TP for TP/TP} and \textbf{EP for DP/EP}.
The two are separated by a load-dependent boundary that production workloads must cross at runtime.

\begin{figure}[t]
    \centering
    \setlength{\fboxsep}{6pt}
    \includegraphics[width=\columnwidth]{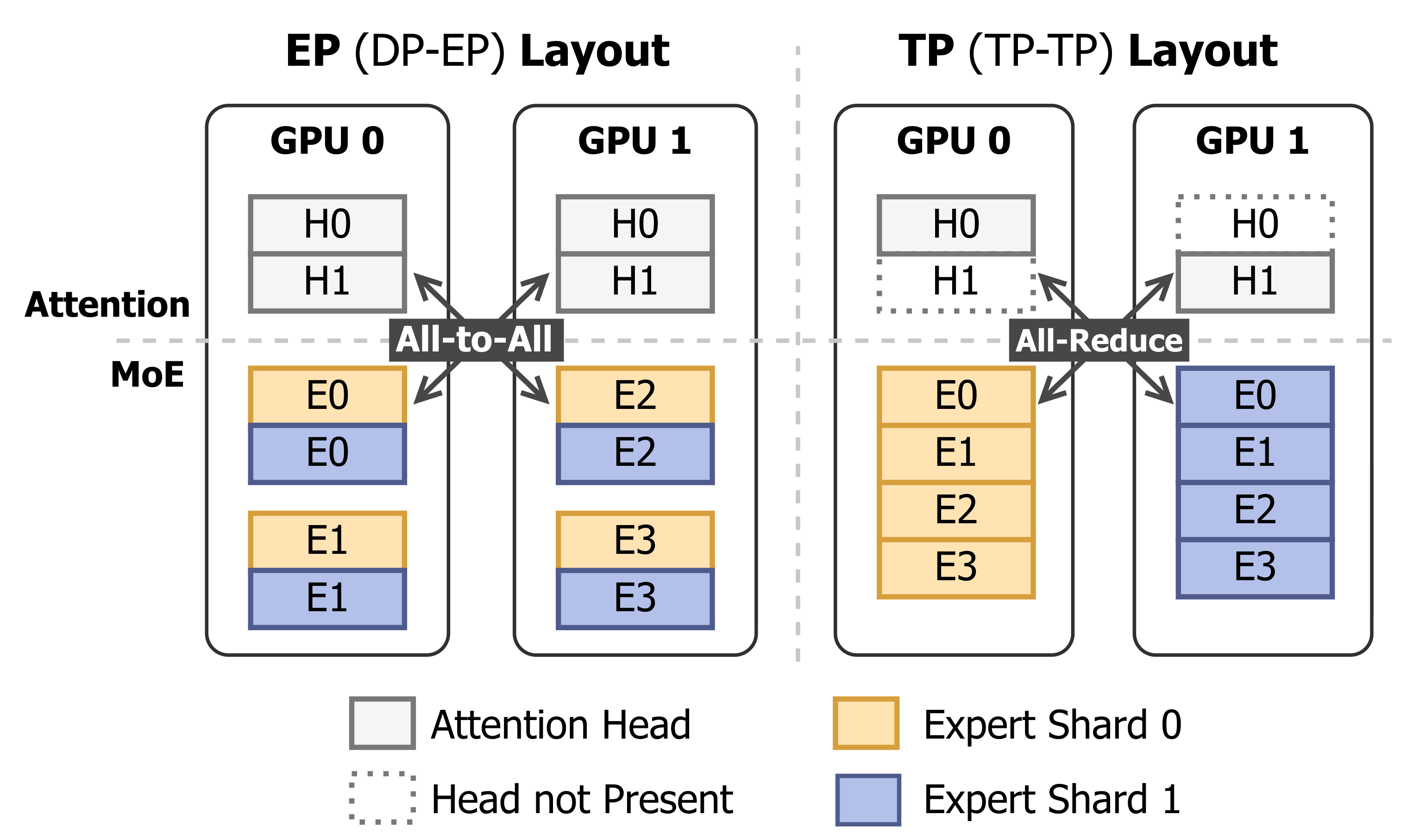}
    \caption{Per-layer layout contrast for \textsc{EP} and \textsc{TP}.
    \textsc{EP} runs data-parallel attention and keeps each whole expert on one rank;
    \textsc{TP} shards both attention heads and individual experts across ranks.}
    \label{fig:bg-layouts}
\end{figure}

\parab{Why the boundary exists.}
\textsc{TP} and \textsc{EP} differ along two axes, each of which flips direction as \(B\) grows (\figref{fig:bg-layouts}).
The first is \emph{communication}: \textsc{TP}'s per-layer \texttt{All-Reduce} ships the full hidden state and grows with \(B\), while \textsc{EP}'s \texttt{All-to-All} carries only routed tokens but pays a small-message dispatch floor that dominates when \(B\) is low.
The second is \emph{MoE compute}: decode MoE GEMMs are memory-bound, so per-rank runtime tracks per-rank token count, which is \(B\) under \textsc{TP} and \(B/G\) under \textsc{EP}, where \(G\) is the EP group size.
Both axes favor \textsc{TP} at small \(B\) and \textsc{EP} at large \(B\), and the boundary sits where their crossovers compound.

\parab{The crossover is hardware- and model-agnostic.}
Both axes are structural, not artifacts of our model and GPU count: \texttt{All-Reduce} volume and the memory-bound \(B\) versus \(B/G\) compute gap hold for any MoE on any interconnect.
Public benchmarks spanning many models and GPU generations show the same latency-throughput frontier, its tight-latency end served by \textsc{TP} and its high-throughput end by \textsc{EP}~\cite{semianalysis_inferencex}.

\subsection{Real World Workloads Cross the Boundary}
\label{sec:bg-workloads}

Production workloads do not sit on one side of the TP--EP boundary.
Two dominant deployment scenarios, bursty online serving and reinforcement-learning (RL) rollouts, drive the active batch across it.

\parab{Bursty online serving.}
Production LLM traces exhibit bursty arrivals separated by long quiet periods~\cite{azurellmtrace, splitwise}, with peak in-flight counts spanning two to three orders of magnitude over minutes.
A single deployment therefore crosses the boundary repeatedly.
EP sustains the throughput required to drain bursts that would saturate TP, while TP delivers the per-token latency that interactive serving demands during quiet periods.

\parab{RL rollouts.}
RL post-training algorithms such as GRPO~\cite{grpo} and DAPO~\cite{dapo} interleave policy updates with rollout steps, and each step is itself a serving workload.
A step submits a batch of prompts, samples each prompt multiple times for group-relative advantage, and decodes every sample to completion before the next weights update, so the engine sees thousands of in-flight requests at the burst peak.
Per-request output lengths vary by more than an order of magnitude across the batch, since reasoning chains terminate at different points and a small fraction run all the way to the multi-thousand-token cap~\cite{rollpacker,XiaomiR3,dapo,laminar,longcatdora}.
The active batch therefore starts well above the boundary, decays through it, and lingers in a long tail of stragglers.

Different RL framework designs handle this tail differently, but none lift the burden from the generation engine.
Synchronous (on-policy) frameworks~\cite{hybridflow,hu2024openrlhf,rlhfuse} wait for the full tail before each weights update, since on-policy training requires every sample to come from the current weights and so accepts the long-tail cost as the price of training on its own outputs.
Asynchronous (off-policy) systems~\cite{streamrl,areal,laminar} relax this constraint by overlapping the tail of one step with training on the previous step, reclaiming GPU cycles but leaving each rollout step's workload shape unchanged, so the generation engine still observes the same burst-to-tail decay.
Partial rollout, another off-policy approach, sidesteps the tail by truncating long generations and resuming them in a later step~\cite{kimik1.5,areal}, but a single response then spans multiple policy versions, which slows convergence in practice~\cite{laminar}.
Outside the partial-rollout regime, every step's in-flight count crosses the TP--EP boundary mid-step, and the generation engine benefits from following the favorable layout regardless of the surrounding framework.

\subsection{Prior Switching Does Not Transfer to MoE}
\label{sec:bg-prior-switching}

The workloads in \secref{sec:bg-workloads} call for an engine that switches between \textsc{TP} and \textsc{EP}.
Runtime switching is not new: dense-model systems already reconfigure layouts under shifting load (Shift Parallelism~\cite{shiftparallelism}, Amoeba~\cite{amoeba}, Flying Serving~\cite{flyingserving}).
Their designs assume what MoE denies: that weights stay \emph{resident}, that the engine can be \emph{drained}, and that the runtime is \emph{free} to rebuild.
A live MoE switch breaks all three.

Weights does not stay resident: a dense TP shard is a $1/P$ slice of the data-parallel replica, so a system that keeps the replica already owns every TP shard without a reshard~\cite{amoeba,flyingserving,shiftparallelism}, but MoE experts dominate the model and EP and TP partition them along orthogonal axes that share no bytes, so a switch must reshard experts across GPUs without holding both copies.
The engine cannot be drained: prior systems let in-flight requests finish or discard them~\cite{shiftparallelism,flyingserving}, whereas a live switch must carry the decode batch and its paged KV cache across EP's and TP's attention layouts without losing work.
The runtime is expensive to rebuild: each layout's attention metadata~\cite{flashinfer,flashattention}, communication buffers~\cite{DeepseekAIDeepEP, ucclep, PerplexityAIpplx-kernels}, and CUDA graphs take tens of seconds to rebuild and stall every request, so a live switch keeps them resident instead.

\section{Adaptive Parallelism}
\label{sec:adaptive-parallelism}

Decode batch size shifts as requests arrive and complete, so neither
mode stays efficient for long. \sysname therefore treats the parallelism mode as
runtime-reconfigurable state rather than a deployment-time constant.
We write EP\(\to\)TP and TP\(\to\)EP for the two switch directions and EP\(\leftrightarrow\)TP when the direction does not matter.
The key
observation is that an EP\(\leftrightarrow\)TP switch only changes ownership: the model weights, in-flight requests, and KV values are the same
before and after the switch across the all ranks. What changes is which rank owns each slice and which mode-specific view each rank uses for subsequent steps.

\subsection{Weights Resharding}
\label{sec:weights-resharding}

\begin{figure}[t]
	\centering
	\includegraphics[width=\columnwidth]{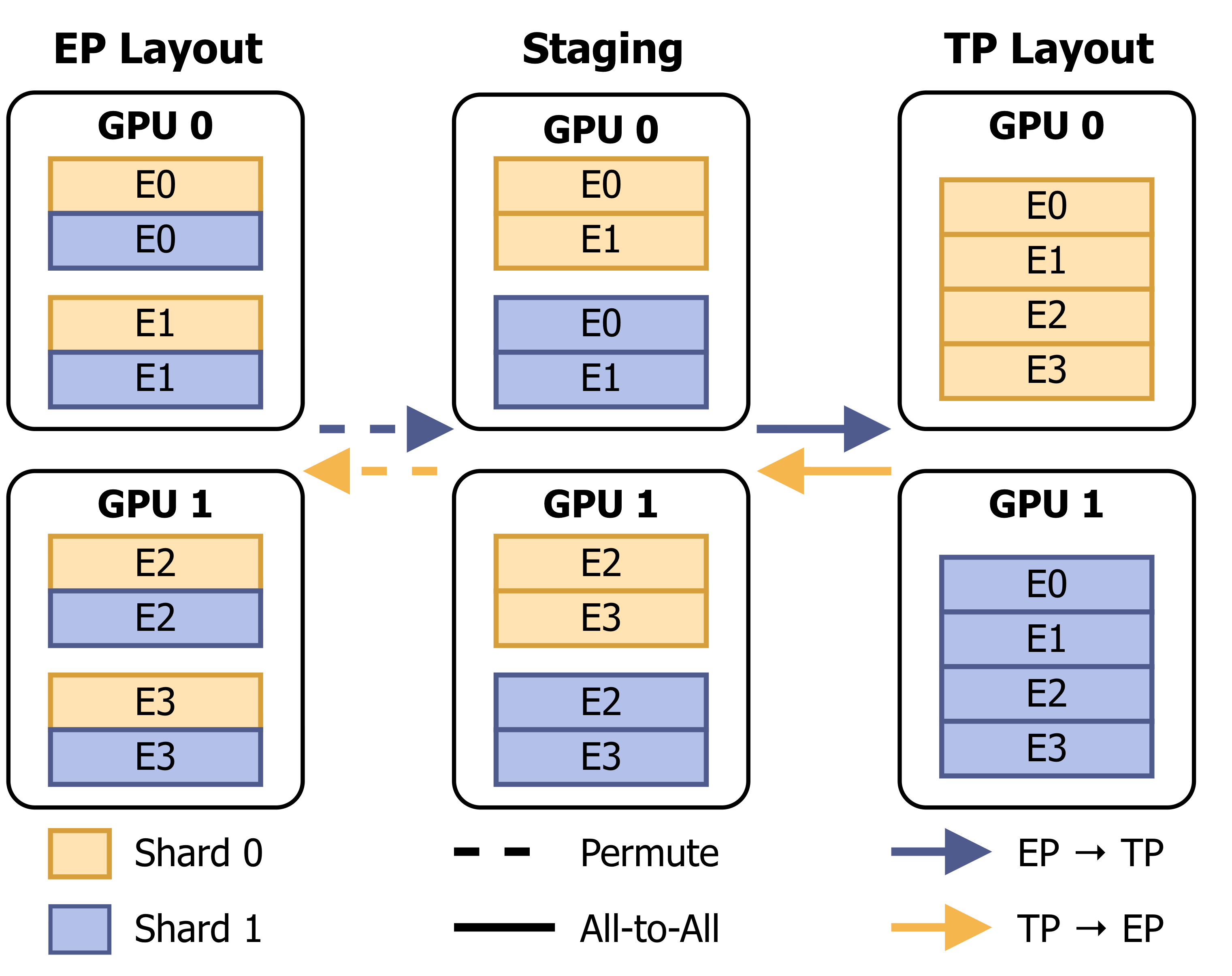}
	\caption{Expert-weight resharding for a switch. EP\(\to\)TP packs local experts into per-peer chunks before one \texttt{All-to-All}; TP\(\to\)EP exchanges data first and then reconstructs complete experts locally.}
	\label{fig:weights-resharding}
\end{figure}

Let \(E\) be the total number of experts, \(P\) the number of ranks in the
switching group, \(H\) the hidden size, and \(I\) the expert intermediate size.
EP and TP store the same global expert weights but assign ownership differently.
Under EP, each rank owns a disjoint subset of complete experts. It owns
\(E_\text{local}=E/P\) experts and stores the full gate-up projection
\(W_{13}\) with shape \((E_\text{local}, 2I, H)\), where the leading factor of two stacks the gate and up projections of the SwiGLU MLP, and the full down projection
\(W_2\) with shape \((E_\text{local}, H, I)\). Under TP, each rank owns one
shard of every expert: \(W_{13}\) has shape \((E, 2I/P, H)\), and \(W_2\)
has shape \((E, H, I/P)\). A switch therefore changes which rank owns each
slice and which tensor view the forward path uses; it does not change the global
weight values.

A reshard decomposes into three possible stages: a local permute that packs
outbound bytes into one contiguous chunk per destination peer, an \texttt{All-to-All} that
transfers ownership across the global communication group, and a local scatter that writes inbound
bytes into the target layout. Weight resharding needs only two stages in either
direction (\figref{fig:weights-resharding}). EP\(\to\)TP runs permute then
exchange: \sysname packs each rank's complete experts into per-peer chunks, and
the \texttt{All-to-All} delivers each rank its shard of every expert already in place.
TP\(\to\)EP runs exchange then permute: the \texttt{All-to-All} delivers contiguous
expert blocks, and the local permute interleaves the received shards into
complete experts. The gate-up and down projections reshard symmetrically, along
the output and input intermediate dimensions, respectively.

Attention weights are small compared with expert weights, so \sysname handles
them separately. Under EP, attention is data-parallel and every rank holds the
full query, key, value, and output projections. Under TP, attention is
tensor-parallel and each rank holds only its head shard. By default, \sysname
keeps both layouts resident and switches attention weights by pointer swap,
paying one duplicated head shard per rank. A memory-saving variant keeps only
the active layout and rebuilds TP\(\to\)EP with an \texttt{All-Gather}.

\subsection{Request Redistribution}
\label{sec:request-redistribution}

\begin{figure}[t]
	\centering
	\includegraphics[width=\columnwidth]{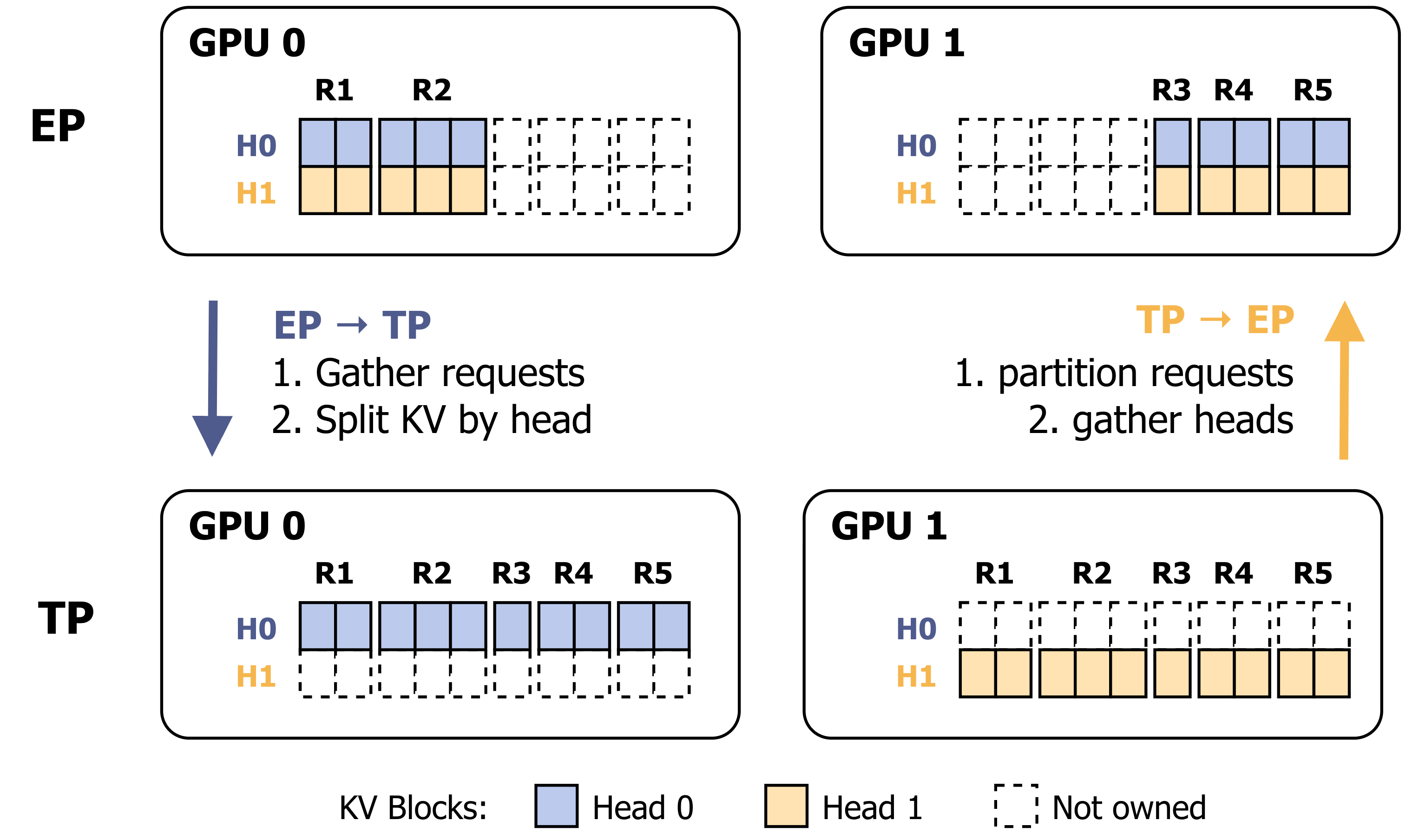}
	\caption{Request and KV-cache redistribution for an EP\(\to\)TP switch. Requests become shared across TP ranks, while paged KV blocks are repartitioned by attention head; TP\(\to\)EP reverses the mapping.  }
	\label{fig:request-redistribution}
\end{figure}

Requests and their KV cache follow the attention layout, so a switch rewrites
ownership of both while preserving the request state and KV values. Under
EP, attention is data-parallel: each rank owns a disjoint subset of in-flight
requests and stores their KV cache for every head. Under TP, attention is
tensor-parallel: every rank serves every request but stores only its slice of the
KV heads (\figref{fig:request-redistribution}). When a model has fewer KV heads
than ranks, TP replicates each head across a group of ranks. For example, Qwen3's four KV heads~\cite{qwen3} at TP~8 place two ranks per head. Request ownership
is only host-resident metadata; the KV cache is large, GPU-resident, and costly
to move.

In the EP\(\to\)TP direction, \sysname first reassigns request ownership
with a metadata \texttt{All-Gather}: each rank contributes its running and waiting
requests, and all ranks construct the same ordered list of in-flight work.
\sysname then repartitions the KV cache by attention head, leaving each rank
with one head shard of every request rather than all heads of its own requests.

Paged attention complicates this transfer. KV cache is stored in fixed-size
blocks drawn from a shared pool, so a request's tokens occupy non-contiguous
slots tracked by a page table, and distinct requests interleave arbitrarily~\cite{vllm}.
Moving the cache request by request would launch thousands of tiny transfers per
layer. \sysname instead reads the page tables and precomputes one index vector
over every token a rank must send. It then gathers the scattered slots into
contiguous per-peer chunks, exchanges those chunks with one \texttt{All-to-All}, and
scatters the received heads into target pages reallocated under TP. Unlike weight resharding, KV
transfer keeps all three stages because paging scatters both the source and
destination slots.

The TP\(\to\)EP direction uses the same cache-transfer stages but changes
request ownership. Instead of gathering requests onto every rank, \sysname
partitions the global request list into disjoint per-rank subsets. It sorts
requests by decreasing sequence length and greedily places each request on the
least-loaded rank, balancing request and token counts together. The heuristic is
deterministic, so every rank computes the same partition without communication.
Each rank then sends its head shard of every departing request to that request's
new owner, which reassembles the full set of heads.

The switch runs synchronously across ranks while decode is paused between
iterations. Request metadata, sampling state, and KV values migrate together, so
no request is dropped and each resumes decoding at the same position. Waiting
requests carry no KV cache, so the switch remaps only their ownership to the
target layout.

\section{System Design}
\label{sec:system}

Building on the switch definition in \secref{sec:adaptive-parallelism},
\sysname must execute each EP\(\leftrightarrow\)TP transition as a low-latency
serving operation rather than an offline reconfiguration. A practical switch
faces four systems challenges, in the order addressed below: avoiding a second
copy of model weights and KV cache, minimizing switch latency for expert-weight
and KV-cache movement, preserving CUDA graphs and other mode-specific runtime
state, and finding a proper point to switch. This section describes how
\sysname addresses these challenges with its memory layout, fused direct-transfer
kernels, runtime preservation, and switch policy.

\subsection{Overview}

\begin{figure}[t]
	\centering
	\includegraphics[width=0.95\columnwidth]{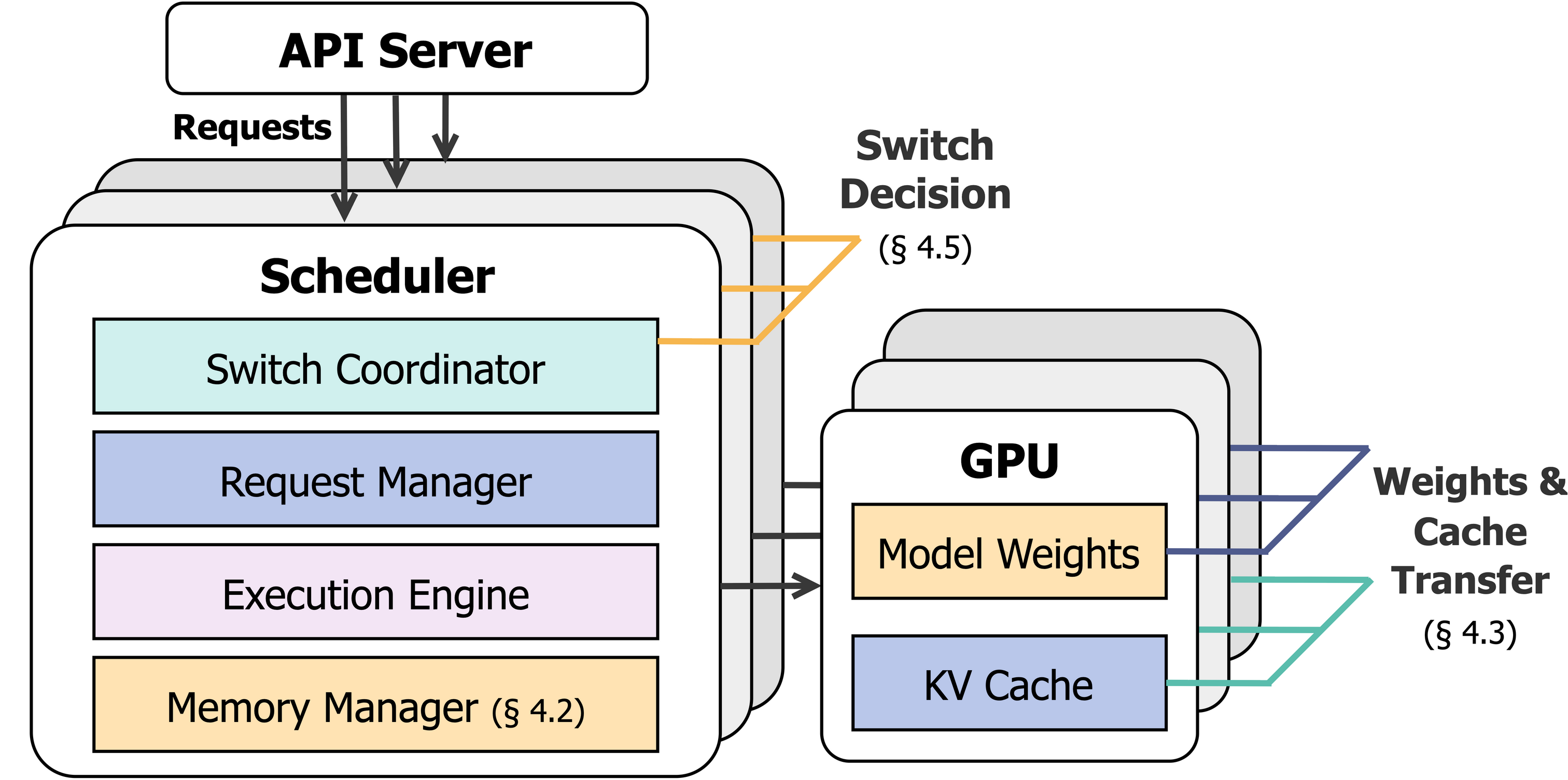}
	\caption{\sysname architecture. Each of stacked boxes denotes one scheduler or one
		GPU rank. The coordinator decides switches
		(\secref{sec:switch-policy}); the GPU holds expert weights and
		KV cache that \sysname reshards over the GPU interconnect without a second copy
		(\secref{sec:direct-data-transfer}).}
	\label{fig:system-overview}
\end{figure}

\figref{fig:system-overview} shows how \sysname embeds switching inside the
serving runtime. A centralized API server admits requests and dispatches them
to per-rank schedulers. Each scheduler contains four components: a switch
coordinator that chooses the target mode, a request manager that tracks request
metadata and KV-cache ownership, an execution engine that runs the active EP or
TP forward path, and a memory manager that owns the rank's GPU-resident weights,
KV cache, and transfer staging buffers.

\sysname separates each switch into a control plane reconfiguration and a
data-plane update. The control plane is small, so \sysname holds both the EP and
TP copies resident and a switch selects prepared state instead of rebuilding it.
The data plane, expert weights and KV-cache pages, is too large to replicate at
model scale, so \sysname reshards the single resident copy over the interconnect.

Switches run between consecutive model forward steps. Before every step, the switch coordinator on rank~0 decides whether to switch modes according to its policy. And the decision is broadcast to other ranks. When a switch is triggered, all ranks enter the transition together: execution engines select the prepared runtime state for the target mode, request managers update request and KV-cache ownership, and memory managers move expert slices and KV pages directly between GPUs. Decoding resumes under the new layout after all ranks complete the transition.

The rest of this section describes the runtime support in dependency order:
the unified memory manager fixes the storage layout
(\secref{sec:static-memory-manager}); direct-transfer kernels fuses all transfer operations (\secref{sec:direct-data-transfer}); the
runtime-preservation keeps CUDA graphs and communication state valid
(\secref{sec:runtime-preserving}); and the switch policy decides when the
transition is worth taking (\secref{sec:switch-policy}).

\subsection{Unified Memory Manager}
\label{sec:static-memory-manager}

\begin{figure}[t]
	\centering
	\includegraphics[width=\columnwidth]{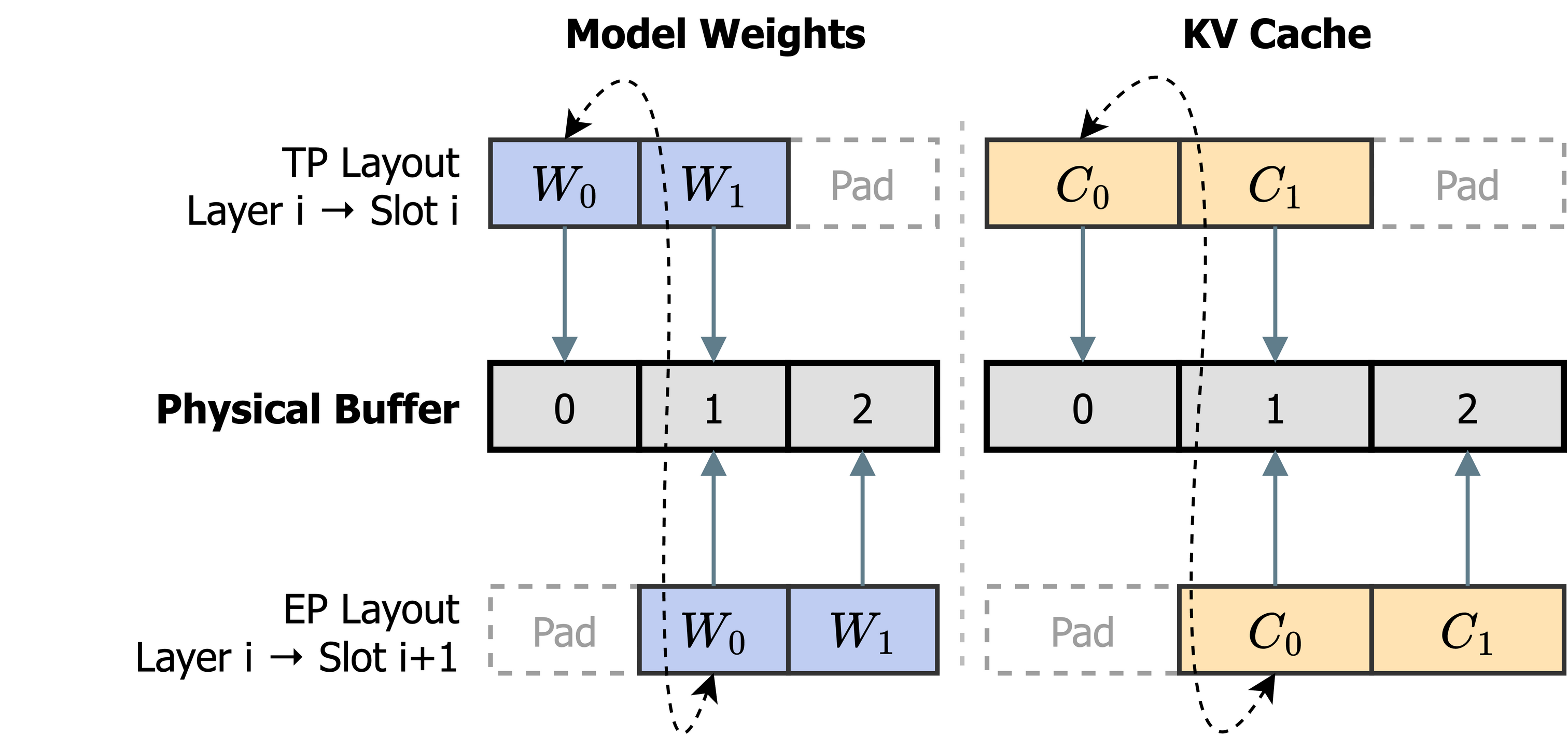}
	\caption{Unified memory manager. Each rank allocates one large GPU buffer and
		serves model weights, KV-cache pages, request buffers, and transfer scratch
		as tensor views into that buffer. For expert weights, \sysname reserves
		\(N{+}1\) slots for \(N\) layers and defines mode-specific aliases: TP maps
		layer \(i\) to slot \(i\), while EP maps layer \(i\) to slot \(i{+}1\).
		The one-slot offset gives each layer distinct source and destination slots
		during resharding.
    }
	\label{fig:memmgr}
\end{figure}

The unified memory manager (UMM) is \sysname's GPU-memory foundation for
switchable states. Instead of letting model components allocate device tensors
independently, each rank allocates one large contiguous unified buffer on GPU at startup, sized by the memory fraction reserved for model weights and KV cache.
UMM then serves all long-lived switch state from this buffer, including expert weights, attention weights and KV-cache pages. Each later allocation is replaced with a tensor view into a fixed region of the unified buffer, so the runtime can reason about both the physical address and the mode-specific layout of every object involved in a switch.

UMM builds this layout with slots and aliases. For expert weights, it divides the
buffer into fixed-size per-layer slots and creates two sets of tensor views over
them, one for TP and the other for EP. The two views expose the shape and stride that
the corresponding forward path expects, but they alias the same rank-local
backing buffer. To make in-place resharding safe, UMM reserves one extra slot:
for \(N\) layers, TP maps layer \(i\) to slot \(i\), while EP maps layer \(i\) to
slot \(i{+}1\) (\figref{fig:memmgr}). During a switch, the transfer kernel reads
from the source-mode alias and writes to the target-mode alias. The one-slot
offset gives each layer separate source and destination slots, and the transfer
uses the direction-specific layer order to avoid overwriting a slot before its
old contents have been read. EP \(\rightarrow\) TP adopts the sequential order while TP \(\rightarrow\) EP adopts the reverse order.

This layout gives \sysname three switch-path properties. First, a switch does
not allocate dynamic buffers for new weights or cache pages; target weight slots and 
KV-cache pages already exist inside the unified buffer.
Avoiding dynamic allocation matters because new device allocations can trigger
PyTorch memory garbage collection and add significant delay to the switch.
Second, as the unified buffer never moves, each rank exports it once as
CUDA IPC memory, and peer GPUs write target slots directly over NVLink (\secref{sec:direct-data-transfer}). Third,
each mode-specific tensor keeps a fixed device-side memory address. CUDA graphs captured separately for EP and TP therefore remain valid across
switches; a mode's buffers may contain stale bytes while inactive, but the graph
still refers to the same addresses when that mode becomes active again. These
properties give fused transfer kernel and runtime preservation the stable storage
layout they depend on.


\subsection{Fused Direct Transfer Kernel}
\label{sec:direct-data-transfer}

\begin{figure}[t]
	\centering
	\includegraphics[width=\columnwidth]{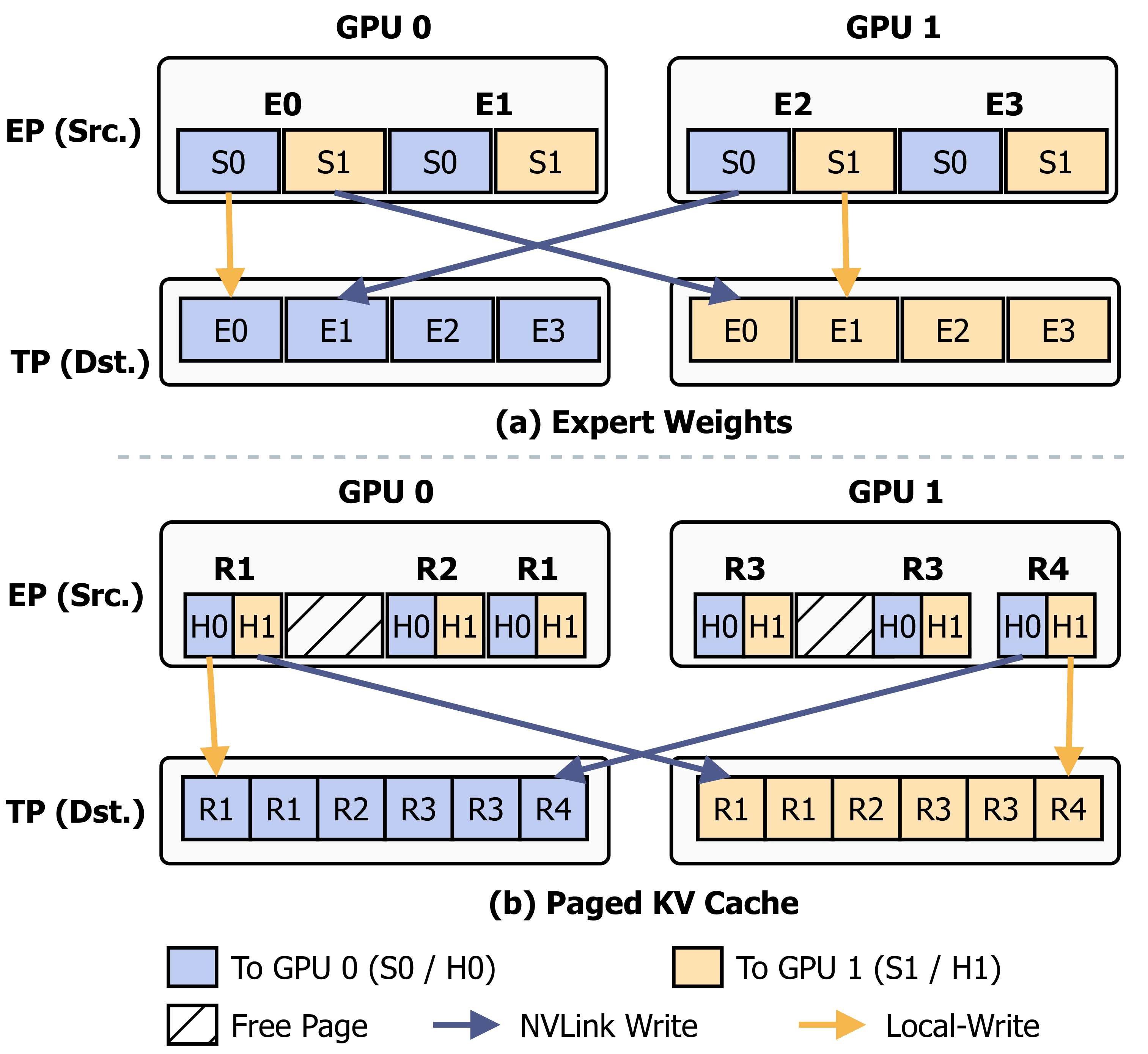}
	\caption{Fused direct-transfer kernels for an EP\(\to\)TP switch; TP\(\to\)EP is symmetric
		with reversed descriptors. Each rank writes its \textbf{(a)} expert-weight shards and
		\textbf{(b)} paged-KV slices straight into the destination slot with no staging buffer or
		\texttt{All-to-All}. Inter-GPU traffic is shown in blue while on-device copy shown in orange. In (b), each token holds two heads with head \(H_k\) routed to rank~\(k\).}
	\label{fig:direct-access}
\end{figure}

The resharding plans of \secref{sec:weights-resharding} and
\secref{sec:request-redistribution} share a common process: gather source bytes into
per-peer chunks, exchange them across ranks, and scatter the received bytes into
the target layout. The straightforward implementation maps this plan to NCCL
collectives, staging each layer's per-peer chunks in a buffer and synchronizing
through an \texttt{All-to-All}; the KV cache adds a gather and scatter around the exchange
because paging scatters its source and destination slots.

This NCCL path is correct but slow on the switch critical path. It touches
the same bytes multiple times in HBM, requires staging buffers for each active
layer, and communicates through an \texttt{All-to-All} at every layer. Double-buffering can
pipeline staging for adjacent layers, but it needs a second staging slot and
still moves data through NCCL's internal buffers before the bytes reach their
final slots.

\sysname instead implements switch data movement with a direct transfer CUDA kernel library. UMM gives each rank a fixed backing-buffer address, so each rank
exports that buffer once as a CUDA IPC handle and maps its peers' buffers into
its own address space. The library provides separate kernels for the two
data-plane objects that a switch moves (\figref{fig:direct-access}).
Expert-weight kernels operate on dense expert slices: each kernel reads from the
source-mode alias and writes the slice into the peer's target-mode slot over
NVLink. KV-cache kernels operate on paged cache metadata: they read the page
tables for live requests, read scattered source blocks through those tables, and
write the corresponding target pages directly into peer buffers.

Both kernel families have the same switch-path benefit. They collapse staging,
data exchange, and scatter into a direct write to the final destination, so the switch does not materialize per-peer chunks in intermediate buffers or communicate
through a per-layer \texttt{All-to-All}. The required source and destination regions
already exist inside UMM, so the kernels also avoid dynamic allocation for new
weights or cache pages during the switch.

\tabref{tab:transfer-cost} summarizes the resulting data movement. NCCL stages
every element through HBM: two reads and one write for weights, and an extra read
and write for the cache because scattered pages are first gathered into a
contiguous buffer. The fused direct transfer kernel makes a single pass for both objects: one HBM
read, and one NVLink store into the peer's slot. It allocates no
staging buffer; instead, it utilizes the extra UMM destination slot already needed
for safe in-place resharding. Thus our transfer kernel uses one extra slot like naive
NCCL, while overlap's double buffering costs two. In every method, the \(1/P\) of each object whose destination is the
sending rank itself is a local HBM write rather than an interconnect transfer;
\tabref{tab:transfer-cost} folds this local write into the NVLink column for
simplicity.

\begin{table}[t]
	\centering
	\caption{Per-element HBM and NVLink passes for an EP\(\to\)TP switch, by data
		object; \(S\) is one layer-sized extra slot. Direct uses UMM's destination
		slot rather than a staging buffer.}
	\label{tab:transfer-cost}
	\begin{tabular}{llccc}
		\toprule
		 & Method & Buffer & HBM & NVLink \\
		\midrule
		\multirow{3}{*}{Weights} & Naive   & \(S\)  & 2 + 1 & 1 \\
		                         & Overlap & \(2S\) & 2 + 1 & 1 \\
		                         & Direct  & \(S\)  & 1 + 0 & 1 \\
		\midrule
		\multirow{3}{*}{Cache}   & Naive   & \(2S\)  & 3 + 2 & 1 \\
		                         & Overlap & \(4S\) & 3 + 2 & 1 \\
		                         & Direct  & \(S\)  & 1 + 0 & 1 \\
		\bottomrule
	\end{tabular}
\end{table}

The KV-cache kernels differ from the weight kernels in how they construct this
final-destination write. Expert-weight traffic is dense and balanced by layer,
but KV traffic is fine-grained and depends on the live request set where different
ranks may own different numbers of tokens, and sequence lengths make the
\texttt{All-to-All} volume imbalanced. \sysname therefore builds page-indexed work
descriptors from the current request metadata, as shown in
\figref{fig:direct-access}, and launches KV kernels over those descriptors. The
kernel still performs the same direct write as the weight path, but it derives
its source and destination addresses from the irregular cache layout produced by
paged attention.

\subsection{Runtime Preserving}
\label{sec:runtime-preserving}

A switch changes the data layout, but the serving runtime also depends on
mode-specific execution state. CUDA graphs, communication groups,
EP communication buffers \cite{DeepseekAIDeepEP,ucclep, PerplexityAIpplx-kernels}, and attention-backend metadata \cite{flashinfer,flashattention} must match the active EP or TP forward path and are expensive to reconstruct, yet together they are small enough to keep resident in both modes. \sysname therefore builds both runtime-state bundles at startup and switches between prepared copies at an iteration boundary.

CUDA graphs impose the strongest constraint because graph replay embeds device
memory addresses and forces every input tensor to stay at a fixed address. EP and TP use different forward paths, so \sysname captures one graph
set per mode. Both graph sets reference UMM-managed mode-specific tensors. Across
switches, those tensors keep the same addresses; while a mode is inactive, its
buffers may contain stale bytes, but the next switch writes the target-mode data
back into those same buffers before the graph is replayed. A switch therefore
changes buffer contents, not graph addresses. During startup, \sysname captures
both graph sets against their real layouts, using a weight-only warmup switch to
enter the alternate mode while no request is active.

The same resident-state rule applies to communication and attention state.
\sysname keeps per-mode communication and attention metadata buffers resident.
When the first rank commits a switch, each rank flips the active graph set, communication state, and attention metadata together, and the request manager hands request ownership to the target layout in the same step (\secref{sec:request-redistribution}). The resumed decode step then observes a consistent target-mode runtime without recapturing graphs or rebuilding runtime state.

\subsection{Switch Policy}
\label{sec:switch-policy}

Rank~0 runs the switch coordinator, which decides when the efficiency gained by
changing modes outweighs the switch cost. \sysname uses the global in-flight
request count as the control signal, because it tracks the per-iteration work
that determines the TP--EP crossover. The coordinator
samples this count once per decode iteration, after the current step completes
and before the next step begins, so the policy never interrupts a forward pass.

The policy is asymmetric. When \sysname runs in TP, a sudden load increase can
make TP throughput-bound, so the coordinator switches to EP as soon as the latest
count exceeds a high threshold \(T_h\). When \sysname runs in EP, a temporary dip
below the crossover is less urgent, since switching too early can cause oscillation.
The coordinator therefore switches back to TP only when the mean count over the
last \(W\) iterations falls below a low threshold \(T_\ell\). The band
\(T_\ell \le T_h\) provides hysteresis, and a cooldown \(C\) after every switch
bounds the maximum switching rate.

\sysname sets the thresholds after startup calibration. Once CUDA graphs have
been captured for both modes, the coordinator probes EP and TP decode cost over a
small set of batch sizes and chooses the crossover as the initial threshold.
Interactive serving uses a wider band and a longer averaging window, making EP
sticky during short load dips. Synchronous rollout uses \(T_\ell=T_h\) and
\(W=1\), because the workload arrives as one burst and then drains monotonically.

Before committing a decision, \sysname checks that the target mode has enough KV
capacity for the current live requests. This matters when the model has fewer KV
heads than ranks: TP replicates each head across ranks, reducing aggregate KV
capacity compared with EP (Qwen3-235B has four KV heads on eight ranks). If the
target mode cannot fit the current request and token set, rank~0 cancels the
switch and retries after the cooldown. Otherwise, it broadcasts the target mode,
and all ranks transition together at the next iteration boundary.


\section{Implementation}
\label{sec:implementation}

\sysname is built on SGLang v0.5.5~\cite{sglang}, and its mechanism is a self-contained module of roughly 7{,}400 lines: the unified memory manager, fused direct-transfer kernels, and switch coordinator.
Integrating it into the serving engine modifies only about 200 lines of existing SGLang code, adding new code paths rather than rewriting existing ones.
An operator enables \sysname and sets its policy entirely through launch arguments: the high threshold, hysteresis band, window, and cooldown are command-line parameters, with the crossover auto-calibrated at startup (\secref{sec:switch-policy}).
Running it needs no change to model weights, the request API, or the surrounding training and serving stack, and the switch coordinator runs on rank~0 inside the existing engine process, so there is no separate controller to provision or monitor.

\section{Evaluation}
\label{sec:evaluation}

Our evaluation answers five questions.
(1)~Does adaptive EP\(\leftrightarrow\)TP switching beat the better static layout when load crosses the crossover mid-run?
(2)~Does the win hold for both online serving and batch generation?
(3)~What does a single switch cost?
(4)~What does preserving CUDA graphs across switches save?
(5)~How much memory does keeping both layouts resident add?
We answer (1) and (2) on two workloads that cross the crossover from opposite directions, bursty serving (\secref{sec:eval-bursty}) and RL rollout (\secref{sec:eval-rollout}), then isolate the switch (\secref{sec:eval-switch}), the CUDA-graph cost it avoids (\secref{sec:eval-cudagraph}), and the memory of holding both layouts (\secref{sec:eval-memory}).

\subsection{Experimental Setup}
\label{sec:eval-setup}

\parab{Hardware and model.}
We evaluate \sysname on a single node of 8 NVIDIA H200 GPUs (141\,GB HBM each), fully connected over NVLink, serving the instruction-tuned Qwen3-235B-A22B model in BF16 (235B parameters, 94 layers, 64 query / 4 KV heads).

\parab{Systems.}
We compare three SGLang deployments (\secref{sec:implementation}) that differ only in parallelism layout.
The two static layouts are the production-grade configurations an operator would actually deploy, and bracket the EP\(\leftrightarrow\)TP tradeoff:
\begin{itemize}
	\item \textsc{TP} runs 8-way tensor parallelism for both attention and the MoE experts.
	\item \textsc{EP} runs data-parallel attention with 8-way expert parallelism, using DeepEP for expert dispatch and combine.
	\item \sysname runs both layouts and switches between them at runtime (\secref{sec:switch-policy}): it enters EP when the latest in-flight count exceeds a high threshold \(T_h\), and returns to TP when the count averaged over a window of \(W\) iterations falls below a low threshold \(T_\ell\), with a cooldown \(C\) between switches.
\end{itemize}

\parab{Configuration.}
All three systems share the same configuration: radix cache disabled, overlap scheduling enabled, a 2{,}048-request concurrency cap, 0.85 static memory fraction, and CUDA-graph enabled.
The prefill token cap follows the active layout, sized to avoid OOM under each layout: 8{,}192 tokens in TP and 4{,}096 tokens per rank in EP.

\sysname's switch policy (\secref{sec:switch-policy}) fixes \(T_h=256\) and cooldown \(C=5\,\)s, widening the band and window for interactive serving (\(T_\ell=0.8\,T_h\), \(W=8\)) and collapsing them for rollout (\(T_\ell=T_h\), \(W=1\)).
All runs report output throughput (decode tokens/s) and sample the active parallelism mode and running-request count at 1\,Hz.
Workload-specific metrics are defined per subsection.

\subsection{Bursty Online Serving}
\label{sec:eval-bursty}

\begin{figure}[t]
	\centering
    \includegraphics[width=\columnwidth]{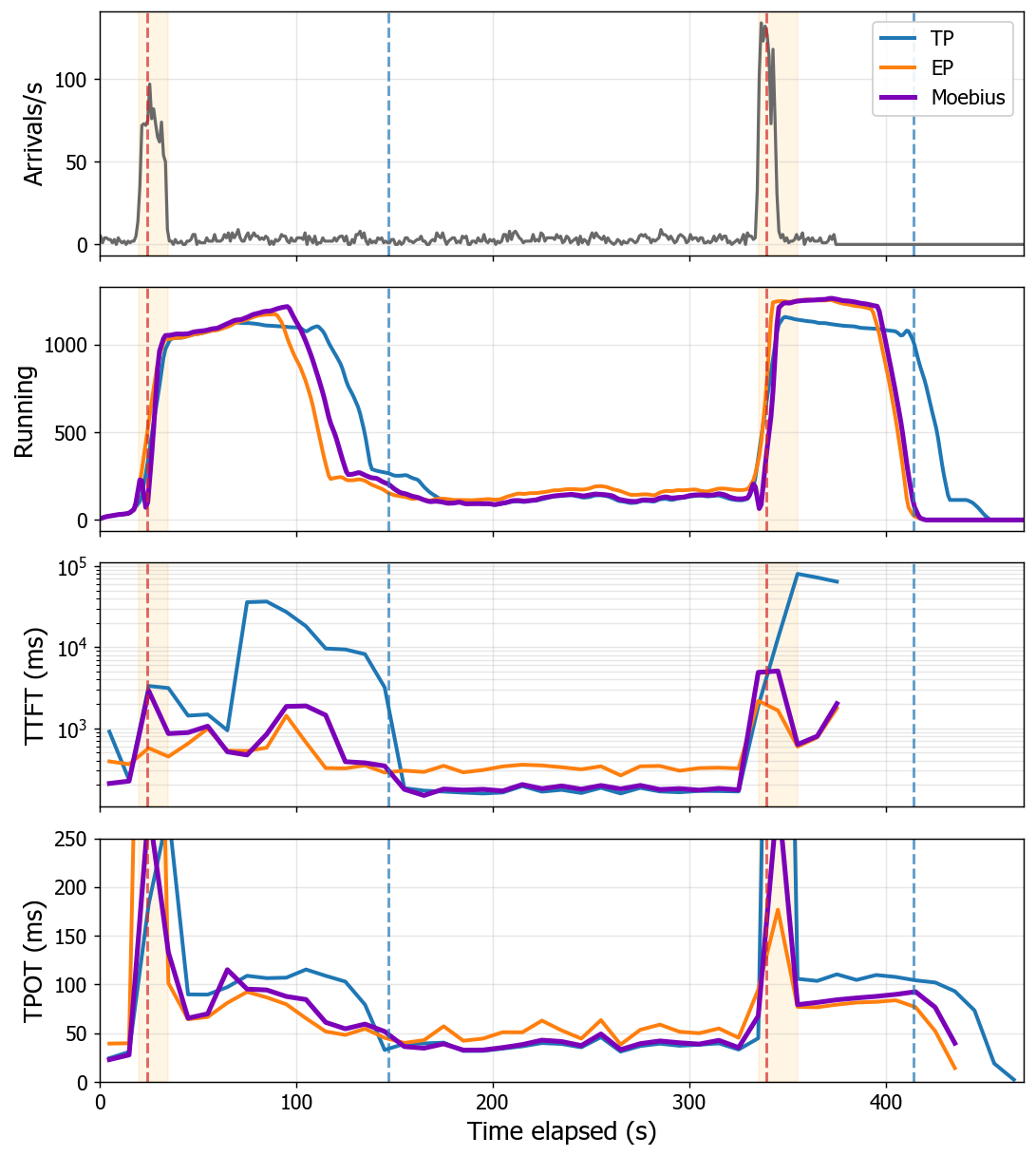}
	\caption{Bursty online serving. Top to bottom: arrival rate, running requests, mean TTFT, and mean TPOT. Orange bands mark burst windows. Dashed lines mark \sysname's TP\(\to\)EP (red) and EP\(\to\)TP (blue) switches.}
	\label{fig:eval-bursty}
\end{figure}

Online request rates rise and fall, so a single deployment crosses the EP\(\leftrightarrow\)TP operating point repeatedly within one run and pays in whichever regime it is not built for.
\sysname instead tracks the favorable mode as the rate moves.

\parab{Workload.}
We replay a 3{,}107-request arrival trace generated by the vLLM bursty workload generator~\cite{vllm}, spanning 375\,s and identical across all three systems.
Two short bursts (peak 80 and 120 requests/s) bracket a 300\,s quiet period at 1--5 requests/s.
Prompts are ShareGPT samples of 300--700 tokens with outputs from \(U(800, 1200)\) tokens.
The bursts drive concurrency well above \(T_h\), and the quiet period drops it back below.
\sysname uses the interactive setting, so it retreats to TP only under sustained low load.
It switches four times over the trace, into EP at each burst onset and back to TP after each burst drains.
We report mean time-to-first-token (TTFT) over arrivals in 10\,s bins and mean time-per-output-token (TPOT) over tokens emitted in the same bins.

\begin{figure}[t]
	\centering
    \includegraphics[width=\columnwidth]{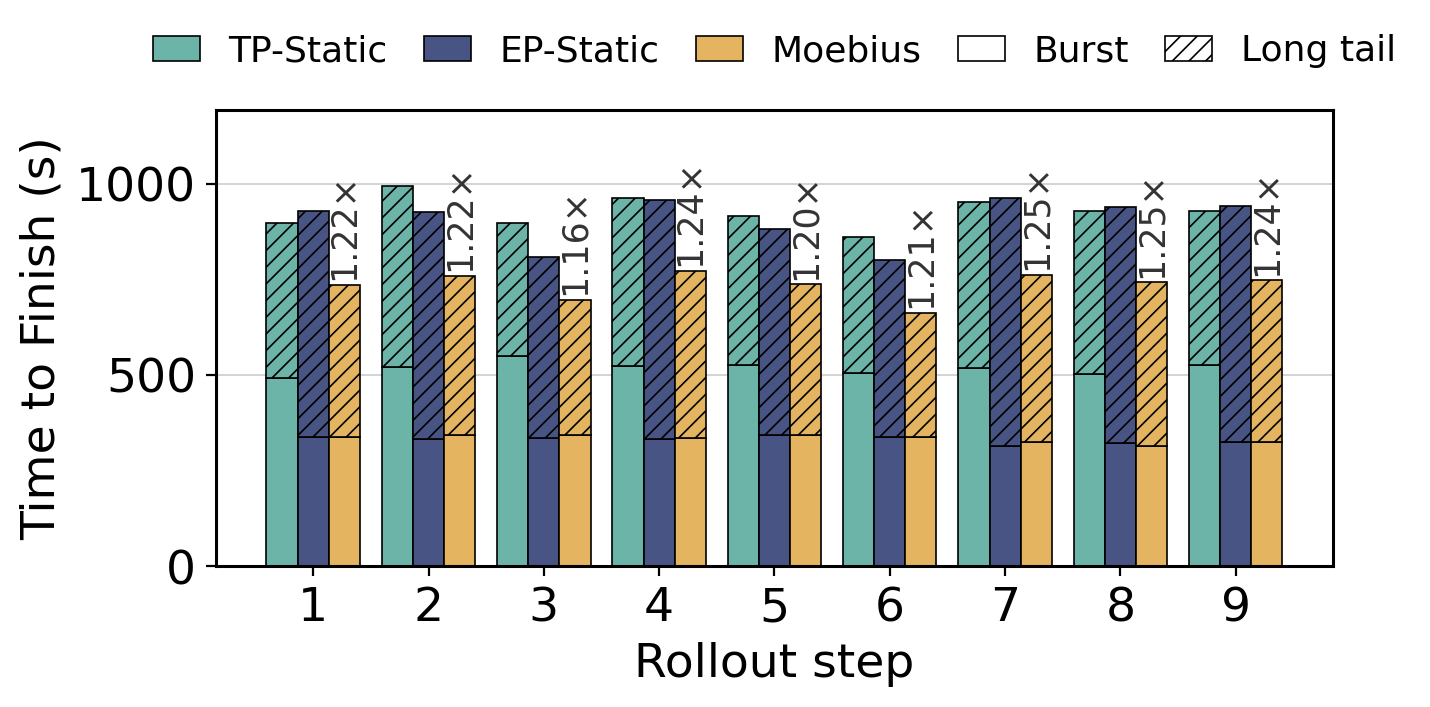}
	\caption{End-to-end completion time for nine DeepMath rollout steps under fixed \textsc{TP}, fixed \textsc{EP}, and \sysname. Each bar is split at the \(T_h=256\) switch threshold into a burst phase (solid) and a long-tail phase (hatched). Labels above \sysname bars show speedup over the better static layout.}
	\label{fig:eval-rollout-e2e-latency}
\end{figure}

\parab{Each static layout pays in one regime.}
Static layouts trade strengths on opposite ends of the trace (\figref{fig:eval-bursty}).
\textsc{TP}'s decode trails \textsc{EP} under burst load.
Its queue outruns it and mean TTFT reaches 9.9\,s, roughly five times \textsc{EP}'s 2.0\,s.
\textsc{EP} absorbs both bursts but pays in the quiet stretch, where \texttt{All-to-All} dispatch overhead lifts TPOT to 52\,ms against \textsc{TP}'s 37\,ms, a 40\% gap that compounds across every decoded token.

\parab{\sysname tracks the favorable layout.}
Running EP through each burst and TP through the quiet period, \sysname stays on the side that is winning at each point of the trace (\figref{fig:eval-bursty}).
Its quiet-period TPOT tracks \textsc{TP} and stays well below \textsc{EP}.
Its burst TTFT peaks at 3.1\,s, three times below \textsc{TP} and close to \textsc{EP}.
On the tail, its p99 TTFT holds at 6.0\,s and 1.9\,s across the two bursts, far below static \textsc{TP}'s 90\,s burst-onset collapse, so the sub-second switch pause never drives the tail.
The residual gap to \textsc{EP} reflects the early-burst TTFT \sysname accrues before its TP\(\to\)EP switch completes (\secref{sec:eval-switch}), the cost of starting each burst in TP.

\subsection{Rollout Workloads}
\label{sec:eval-rollout}

\begin{figure*}[t]
	\centering
	\includegraphics[width=\textwidth]{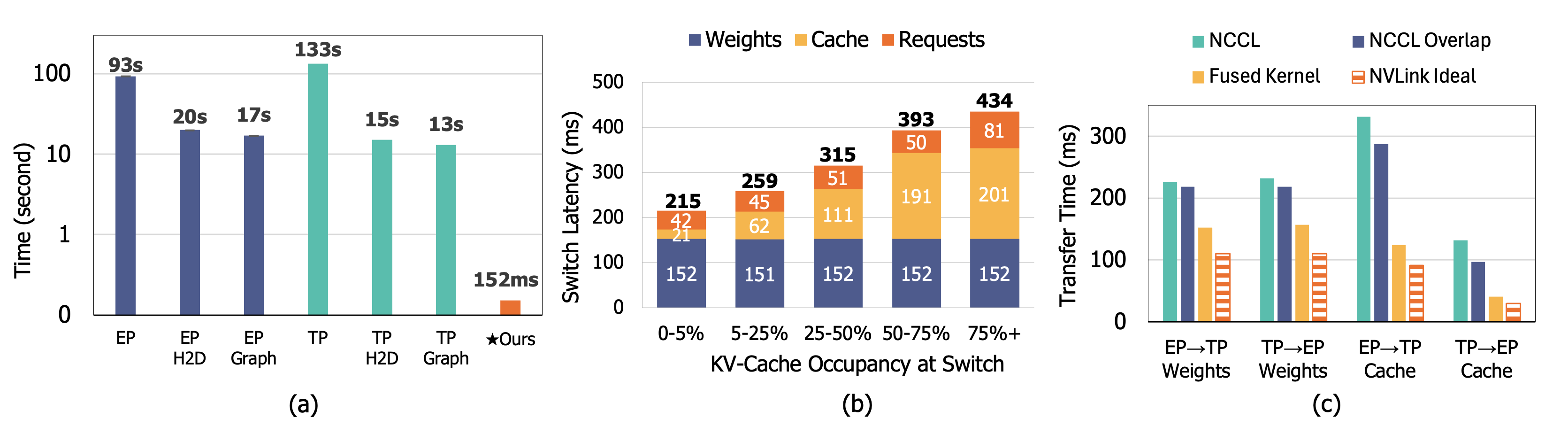}
	\caption{\sysname's switch cost and optimizations. \textbf{(a)} End-to-end switch latency for three strawmen (restart, host-memory weight load, and CUDA-graph recapture) and \sysname's switch, in both directions with the in-flight batch drained. \textbf{(b)} A production EP\(\to\)TP switch decomposed into weight, KV-cache, and request phases, binned by KV-cache occupancy. \textbf{(c)} Transfer time breakdowns for expert weights and the KV cache in both directions, across the fused kernel, NCCL, NCCL with stream overlap, and the NVLink bandwidth ceiling.}
	\label{fig:ablations}
\end{figure*}

Reinforcement-learning post-training, such as GRPO~\cite{grpo} and DAPO~\cite{dapo}, interleaves policy updates with \emph{rollout steps}.
Each step is itself a serving workload: it submits a large batch of prompts and generates every one to completion before the next weights update.
Because output lengths are heavy-tailed, the batch starts large and decays toward a long tail of still-decoding requests.
A rollout step therefore crosses the EP\(\leftrightarrow\)TP operating point \emph{within a single step}: the large initial batch favors EP, the small late-stage batch favors TP.
Any fixed deployment commits to one side of this tradeoff for the entire run.
\sysname instead preserves in-flight requests and changes the parallelism layout as the active batch shrinks.

\parab{Workload.}
We use the DeepMath math-reasoning benchmark~\cite{deepmath} on Qwen3-235B: each rollout step submits a batch of \(N=2048\) prompts and decodes every one to a 32{,}768-token cap.
This is representative of production RL, which runs steps of one to eight thousand prompts under 16k--32k-token caps~\cite{rollpacker,XiaomiR3,dapo,laminar,longcatdora}.
Pooled over the nine steps (18{,}432 requests), inputs are short and clustered (median 120 tokens, max 1{,}352).
Outputs are long and heavy-tailed (median 1{,}510 tokens, p99 10{,}386), with the longest reasoning chains running to the 32k cap (\figref{fig:eval-rollout-dist}).
This asymmetry makes the step decode-dominated and gives the active batch its slow decay: the median request finishes early, while a few long outputs keep decoding an order of magnitude longer and drag the in-flight count down a long tail.

\parab{Methodology.}
To hold the decode work identical across systems, we capture each request's output length once under EP's natural generation, then replay those exact lengths under \textsc{TP} and \sysname.
Every system therefore decodes the same number of tokens for every request, isolating system performance from layout-induced differences in output length.
EP's capture run serves as its own datapoint.
Steps vary in prompt mix and so in output-tail shape, so we evaluate nine steps spanning light to heavy tails, holding the policy fixed so the only variation is the workload.
Because the active batch only shrinks, \sysname uses the rollout setting: it starts in EP and switches once to TP when the batch drains below \(T_h\).
We report end-to-end completion time.

\parab{The win is phase-matching, not a per-token speedup.}
\figref{fig:eval-rollout-e2e-latency} shows \sysname fastest at all nine steps, beating the better static layout by 1.16--1.25\(\times\) (mean 1.22\(\times\)) and the worse by up to 1.31\(\times\).
The more telling result is which static layout is better: \textsc{EP} wins most steps on burst throughput, but on the heaviest-tailed steps, where the rollout lingers in the small-batch tail, \textsc{TP} pulls ahead, so which layout wins depends on the step's prompt mix and is not known until the step has run.
The ``better static layout'' that \sysname is compared against is therefore an oracle that picks the faster of \textsc{TP} and \textsc{EP} per step with foreknowledge, a choice no operator can deploy, yet \sysname beats even this oracle because its gain comes from switching \emph{within} a step rather than committing to one layout for the whole step.
The split bars expose why \sysname beats both: within every step \textsc{EP} clears the burst phase faster while \textsc{TP} decodes the long tail faster, and \sysname runs the burst in EP and the tail in TP, tracking the faster layout in \emph{both} phases.

\parab{End-to-end projection.}
The per-step speedup translates to end-to-end training time through the rollout fraction and the framework's overlap scheme (\secref{sec:bg-workloads}).
For synchronous on-policy training at a 65--85\% rollout fraction, Amdahl's law projects our 1.16--1.25\(\times\) per-step gain to 1.10--1.20\(\times\) end-to-end, and rollout-bound asynchronous schemes pass it through more directly (\appref{app:e2e-projection}).

\subsection{Switch Cost and Optimizations}
\label{sec:eval-switch}

A switch redistributes three pieces of resident state: expert weights, the paged KV cache, and request ownership.
\figref{fig:ablations} isolates its cost three ways.
Panel (a) sets the bar against reconfiguration strawmen that drain the batch and rebuild the target layout, panel (b) decomposes a production EP\(\to\)TP switch by phase across KV-cache occupancy, and panel (c) pits the fused-kernel transfer against NCCL collectives over both components and directions.

\parab{Switching without rebuilding wins by orders of magnitude.}
Panel (a) drains the in-flight batch and compares \sysname's bidirectional switch against three rebuild strawmen (\figref{fig:ablations}(a)).
The strawmen strip one cost at a time: restart pays cold model load and graph recapture, host-memory weight loading drops the disk load, and reusing \sysname's fused transfer leaves only recapture, needed because the rebuilt buffers land at fresh addresses.
Every rung still costs seconds to minutes.
By transferring into pinned addresses and keeping both graph sets resident, \sysname avoids reload and recapture alike, switching in 152\,ms, orders of magnitude faster (\secref{sec:eval-cudagraph}).

\parab{The cheap switch is what makes within-episode adaptivity viable.}
A switch is worth taking only when it costs less than the work it saves.
On the 375\,s bursty trace \sysname switches four times.
At the restart strawman's 93--133\,s per switch, or even the host-reload and graph-recapture variants' 13--20\,s, the switch cost alone would dwarf the trace and make naive adaptivity net-negative.
Beyond cost, every rebuild path must drain the in-flight batch before it can reshard.
Applying one during serving would stall or drop live requests, which production serving cannot accept regardless of latency.
\sysname's switch sits two to three orders of magnitude below this bar while retaining the ongoing requests, which is what lets it follow the favorable layout within an episode rather than only across deployments.
We therefore characterize these rebuild paths as switch-cost bounds (\figref{fig:ablations}(a)) rather than running them as end-to-end baselines.
A full-workload comparison would only restate the loss which their per-switch cost and forced draining already guarantee.

\parab{A switch is a fixed weight floor plus a load-dependent KV term.}
We mine 16 EP\(\to\)TP switches from our rollout and serving runs and split each into its three phases, binning by KV-cache occupancy (\figref{fig:ablations}(b)).
Weight transfer is a fixed cost floor, the same reshard panel (a) measures with the batch drained.
Only the KV phase grows monotonically with occupancy, and it sets the gap between a light and a heavy switch.
Request redistribution tracks the in-flight request count rather than per-request cache footprint, so it stays flat.
Even with the cache full the total stays under half a second, and its spread comes almost entirely from KV volume.

\parab{The win extends to the KV cache and the reverse direction.}
Across KV-cache transfer and the TP\(\to\)EP direction (\figref{fig:ablations}(c)), the fused kernel sustains above 70\% of NVLink peak on every bar, reshards expert weights 1.49\(\times\) faster than NCCL in both directions, and beats it by over 2\(\times\) on the cache.
Even the 70\% floor sits closer to the hardware limit than it appears.
Fusing the layout transform into the copy keeps the transfer on the SMs, which in our measurements top out near 77\% of peak rather than the roughly 90\% a dedicated copy engine reaches, a ceiling unavailable here because a copy engine cannot reshard on the fly.
Against that 77\% attainable ceiling, holding above 70\% of peak already puts the fused kernel over 90\% efficiency, so the reshard is effectively saturating the SM transfer path with little headroom left to recover.
Overlapping the collectives across CUDA streams does not close the gap, because on flat NVL8 every GPU pair shares one fabric, leaving the collectives bandwidth-bound so pipelining pays off only on hierarchical interconnects.
Even EP\(\to\)TP, the costlier direction because TP replicates each KV head across two ranks and so doubles the per-rank bytes, stays well below NCCL, so KV transfer is not the bottleneck.

\subsection{Cost of Preserving CUDA Graphs}
\label{sec:eval-cudagraph}

\sysname captures both the EP and TP graph sets at startup and keeps both resident, so a switch swaps the active graph pointer in under a millisecond rather than rebuilding any graph (\secref{sec:runtime-preserving}).
This avoids two costs at once: the per-switch recapture stall already charged to the strawmen in \secref{sec:eval-switch}, and a per-token tax from decoding eagerly with no graph at all.
We measure that per-token tax on the same Qwen3-235B, 8\(\times\)H200 configuration.

\begin{figure}[t]
	\centering
	\includegraphics[width=0.95\columnwidth]{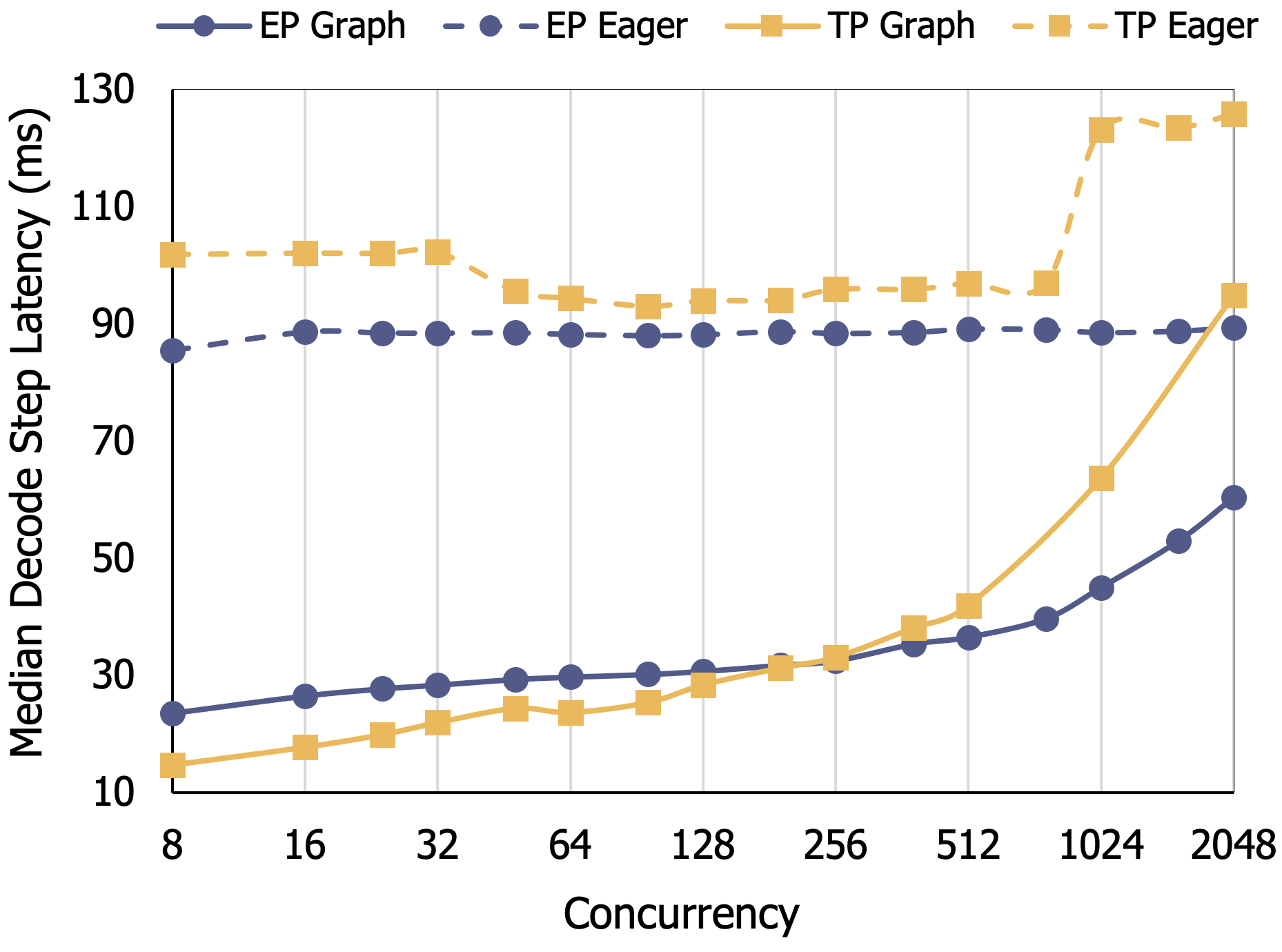}
	\caption{Median per-step decode latency with and without CUDA graphs, across batch sizes.}
	\label{fig:eval-cudagraph}
\end{figure}

\parab{Eager decoding taxes every step.}
Without graphs, each step issues every layer's kernel launches individually instead of replaying them in one graph launch (\figref{fig:eval-cudagraph}).
The host overhead hurts most where compute is smallest, up to \(6.95\times\) at the low batch sizes \sysname runs in TP, and shrinks but never disappears as the batch grows.
That worst case falls in low-concurrency TP, exactly the regime \sysname enters TP to serve, so preserving graphs is what makes that mode viable.
Eager decoding is also unpredictable, with GC and launch-queue stalls spiking occasional steps well above the median.

\parab{Holding both graph sets resident is cheap.}
Each mode caps capture at a per-rank batch of 256 and so holds only 36 graphs, small enough to keep both sets resident (\secref{sec:eval-memory}).
The cap is safe because \sysname enters EP rather than decoding in TP above the \(T_h=256\) crossover, and EP's data-parallel attention keeps each rank at or below 256 even at full concurrency.
Paying this capture once at startup, rather than at every switch, is what turns the per-switch recapture stall of \secref{sec:eval-switch} into a sub-millisecond pointer swap.

\subsection{Memory Footprint}
\label{sec:eval-memory}

\begin{figure}[t]
	\centering
	\includegraphics[width=0.95\columnwidth]{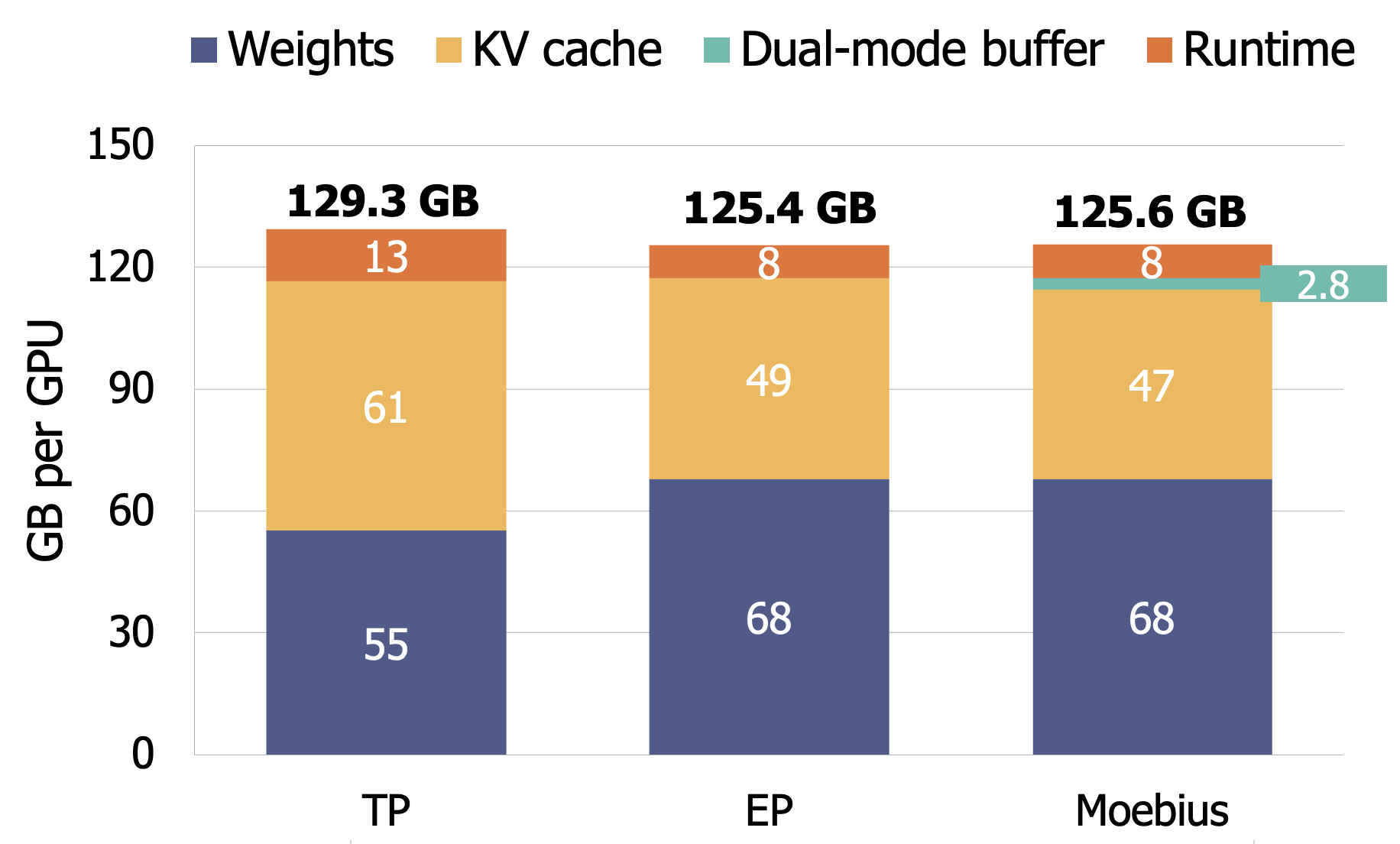}
	\caption{Per-GPU memory footprint at rest, split into weights, KV cache, \sysname's dual-mode buffer, and runtime state.}
	\label{fig:eval-memory}
\end{figure}

\sysname keeps both the EP and TP layouts resident, so a natural concern is that it doubles the memory of a single-layout server.
It does not.
We measure per-GPU memory at rest, after weight load, KV-cache allocation, and CUDA-graph capture but before any request, on the same 8\(\times\)H200, Qwen3-235B, 0.85 memory-fraction configuration as above.
This static snapshot is the whole story: at switch time and during serving \sysname reuses pre-allocated buffers rather than materializing new tensors.
\figref{fig:eval-memory} splits each footprint into model weights, KV cache, \sysname's dual-mode buffer, and runtime state, the last detailed in \appref{app:memory}.

\parab{Two resident layouts, one memory budget.}
Although \sysname holds two CUDA-graph sets and both attention layouts, it fits within \textsc{EP}'s budget: within 0.2\,GB of \textsc{EP} and 3.7\,GB below \textsc{TP} (\figref{fig:eval-memory}).
The heavy state is shared, not duplicated: its weights are identical to \textsc{EP}'s, and the MoE experts, EP communication buffer, and NCCL communicators are allocated once and reused in both layouts.
\sysname and \textsc{EP} carry 12.7\,GB more weight per GPU than \textsc{TP} only because data-parallel attention replicates the attention stack on every GPU, a cost of EP rather than the price of holding two layouts.
Its one genuine overhead is a 2.8\,GB dual-mode buffer: it holds the TP-mode attention shards alongside the full EP copies so a switch stays a pointer swap, plus a spare physical layer that stages the per-layer transfer (\appref{app:memory}).
Because these bytes sit inside the weight allocation, \sysname funds them by forgoing 2.8\,GB of KV cache (\(+2.4\%\) per GPU) rather than adding memory on top.
Even charged the full buffer it stays below \textsc{TP}, since it sizes its TP-mode graphs and workspaces for a batch of 256 rather than \textsc{TP}'s full 2{,}048.

\section{Related Work}
\label{sec:related}

\parab{Runtime parallelism switching.}
Prior runtime switching targets dense serving or stays within training, never a live MoE EP\(\leftrightarrow\)TP serving switch.
SpotServe~\cite{spotserve}, Amoeba~\cite{amoeba}, Flying Serving~\cite{flyingserving}, and Shift Parallelism~\cite{shiftparallelism} switch dense serving among data, tensor, and sequence parallelism, while HotSPa~\cite{hotspa} and Tutel~\cite{MLSYS2023_5616d34c} switch at training-step boundaries.
\sysname reshards its memory-dominant experts without replica, and repartitions the KV cache for in-flight requests.

\parab{Serving and execution for MoE.}
Among MoE systems, HAP~\cite{lin2025haphybridadaptiveparallelism} chooses hybrid layouts offline with no online transitions, while the rest optimize a single fixed layout: DeepSpeed-MoE~\cite{deepspeedmoe} and DeepSpeed-Inference~\cite{10.5555/3571885.3571946} scale expert inference, Lina~\cite{lina} targets communication cost. The fused EP communication kernels~\cite{DeepseekAIDeepEP,PerplexityAIpplx-kernels,ucclep} improve token dispatch, and hybrid mappings~\cite{Singh_2023} still commit to one deployment layout.
\sysname is orthogonal: it switches layouts as load changes and reuses these kernels inside each mode.
It also preserves the standard serving stack across a switch, building on continuous batching (Orca~\cite{orca}), paged KV cache (vLLM~\cite{vllm}), SGLang's EP and TP forward paths~\cite{sglang}, and FlashAttention~\cite{flashattention} and FlashInfer~\cite{flashinfer} kernels, with a unified memory manager that keeps these CUDA graphs valid after a reshard (\secref{sec:static-memory-manager}).

\parab{Serving for RL rollout.}
RL post-training turns rollout into a serving workload.
HybridFlow~\cite{hybridflow} and OpenRLHF~\cite{hu2024openrlhf} orchestrate rollout and training, GRPO~\cite{grpo} and DAPO~\cite{dapo} defines our rollout workload, and StreamRL~\cite{streamrl} and AReaL~\cite{areal} hide rollout cost through disaggregation or asynchrony.
Rollpacker~\cite{rollpacker} and TLT~\cite{TLT} targets eliminating and accelerating the long tail problem. \sysname is orthognal to them and can be integrated in the rollout engine to collectively improve rollout efficiency.

\section{Discussion}
\label{sec:discussion}

\parab{Generality of the mechanism.}
\sysname separates a model-agnostic control path from architecture-specific resharding that depends only on the expert count \(E\), rank count \(P\), hidden size \(H\), and intermediate size \(I\).
Grouped- and multi-head attention, finer-grained or shared experts, and multi-head latent attention therefore fit by changing only how a tensor is sharded, not the control path.
Those speedups are specific to our model and hardware.

\parab{Capacity-aware switching is safe by construction.}
When KV heads are fewer than ranks, TP holds less KV cache, and on eight ranks Qwen3-235B's four KV heads halve TP capacity.
\sysname switches to TP only when the live KV fits and otherwise stays in EP (\secref{sec:switch-policy}), so a canceled switch forgoes a latency gain but never admits an infeasible layout, leaving \sysname bounded below by the better feasible static layout.

\parab{Scope and future work.}
The framework is topology- independent, but the direct-transfer kernel targets a single NVLink domain, where it reshards 1.49\(\times\) faster than an NCCL collective (\secref{sec:eval-switch}).
Across nodes the switch falls back to a collective while still keeping one resident weight copy, preserving CUDA graphs, and migrating live requests, and a direct path over RDMA remains future work.
Our rollout numbers also measure the generation engine alone, so coupling \sysname with a full RL system that spans generation, policy updates, and scheduling, to turn per-step speedups into end-to-end training gains, is likewise left to future work.

\section{Conclusion}
\label{sec:conclusion}

MoE serving shows a TP--EP performance crossover where production workloads cross at runtime. TP decodes small batches with lower latency and EP sustains throughput with large batches.
\sysname makes the layout reconfigurable by treating an EP\(\leftrightarrow\)TP switch as a change of data ownership rather than semantics, resharding resident expert weights and KV cache over the interconnect without a second replica while in-flight requests and captured CUDA graphs survive the switch.
On 8\(\times\)H200 GPUs serving Qwen3-235B, \sysname is fastest on every RL-rollout step, and completes each switch in 215--434\,ms at 2.4\% memory overhead, beating the better static layout by 1.16--1.25\(\times\).

\clearpage

\bibliographystyle{ACM-Reference-Format}
\bibliography{main}

@article{amoeba,
  author = {Haoyu Chen and Xue Li and Kun Qian and Yu Guan and Jin Zhao and Xin Wang},
  title = {{Amoeba: Runtime Tensor Parallel Transformation for LLM Inference Services}},
  journal = {arXiv preprint arXiv:2509.19729},
  year = {2026}
}

@inproceedings{flyingserving,
  author = {Shouwei Gao and Junqi Yin and Feiyi Wang and Wenqian Dong},
  title = {{Flying Serving: On-the-Fly Parallelism Switching for Large Language Model Serving}},
  booktitle = {Proceedings of the 40th ACM International Conference on Supercomputing (ICS)},
  year = {2026}
}

@inproceedings{shiftparallelism,
  author = {Hidayetoglu, Mert and Qiao, Aurick and Wyatt, Michael and Rasley, Jeff and He, Yuxiong and Rajbhandari, Samyam},
  title = {{Shift Parallelism: Low-Latency, High-Throughput LLM Inference for Dynamic Workloads}},
  booktitle = {Proceedings of the 31st International Conference on Architectural Support for Programming Languages and Operating Systems (ASPLOS)},
  year = {2026}
}

@inproceedings{shazeer2017outrageouslylargeneuralnetworks,
  author = {Noam Shazeer and Azalia Mirhoseini and Krzysztof Maziarz and Andy Davis and Quoc Le and Geoffrey Hinton and Jeff Dean},
  title = {{Outrageously Large Neural Networks: The Sparsely-Gated Mixture-of-Experts Layer}},
  booktitle = {Proceedings of the 5th International Conference on Learning Representations (ICLR)},
  year = {2017}
}

@article{megatronTP,
  author = {Mohammad Shoeybi and Mostofa Patwary and Raul Puri and Patrick LeGresley and Jared Casper and Bryan Catanzaro},
  title = {{Megatron-LM: Training Multi-Billion Parameter Language Models Using Model Parallelism}},
  journal = {arXiv preprint arXiv:1909.08053},
  year = {2020}
}

@misc{DeepseekAIDeepEP,
  author = {Chenggang Zhao and Shangyan Zhou and Liyue Zhang and Chengqi Deng and Zhean Xu and Yuxuan Liu and Kuai Yu and Jiashi Li and Liang Zhao},
  title = {{DeepEP: an efficient expert-parallel communication library}},
  year = {2025},
  note = {\url{https://github.com/deepseek-ai/DeepEP}}
}

@misc{PerplexityAIpplx-kernels,
  author = {Perplexity-AI},
  title = {{Efficient and Portable Mixture-of-Experts Communication}},
  year = {2025},
  note = {\url{https://github.com/perplexityai/pplx-kernels}}
}

@inproceedings{dapo,
  author = {Qiying Yu and Zheng Zhang and Ruofei Zhu and Yufeng Yuan and Xiaochen Zuo and Yu Yue and Weinan Dai and Tiantian Fan and Gaohong Liu and Lingjun Liu and Xin Liu and Haibin Lin and Zhiqi Lin and Bole Ma and Guangming Sheng and Yuxuan Tong and Chi Zhang and Mofan Zhang and Wang Zhang and Hang Zhu and Jinhua Zhu and Jiaze Chen and Jiangjie Chen and Chengyi Wang and Hongli Yu and Yuxuan Song and Xiangpeng Wei and Hao Zhou and Jingjing Liu and Wei-Ying Ma and Ya-Qin Zhang and Lin Yan and Mu Qiao and Yonghui Wu and Mingxuan Wang},
  title = {{DAPO: An Open-Source LLM Reinforcement Learning System at Scale}},
  booktitle = {Proceedings of the Advances in Neural Information Processing Systems 38 (NeurIPS)},
  year = {2025}
}

@inproceedings{hu2024openrlhf,
  author = {Jian Hu and Xibin Wu and Wei Shen and Jason Klein Liu and Weixun Wang and Songlin Jiang and Haoran Wang and Hao Chen and Bin Chen and Wenkai Fang and Xianyu and Yu Cao and Haotian Xu and Yiming Liu},
  title = {{OpenRLHF: A Ray-based Easy-to-use, Scalable and High-performance RLHF Framework}},
  booktitle = {Proceedings of the 2025 Conference on Empirical Methods in Natural Language Processing: System Demonstrations (EMNLP)},
  year = {2025}
}

@inproceedings{hybridflow,
  author = {Sheng, Guangming and Zhang, Chi and Ye, Zilingfeng and Wu, Xibin and Zhang, Wang and Zhang, Ru and Peng, Yanghua and Lin, Haibin and Wu, Chuan},
  title = {{HybridFlow: A Flexible and Efficient RLHF Framework}},
  booktitle = {Proceedings of the 20th European Conference on Computer Systems (EuroSys)},
  year = {2025}
}

@inproceedings{deepspeedmoe,
  author = {Samyam Rajbhandari and Conglong Li and Zhewei Yao and Minjia Zhang and Reza Yazdani Aminabadi and Ammar Ahmad Awan and Jeff Rasley and Yuxiong He},
  title = {{DeepSpeed-MoE: Advancing Mixture-of-Experts Inference and Training to Power Next-Generation AI Scale}},
  booktitle = {Proceedings of the 39th International Conference on Machine Learning (ICML)},
  year = {2022}
}

@inproceedings{hotspa,
  author = {Ge, Hao and Fu, Fangcheng and Li, Haoyang and Wang, Xuanyu and Lin, Sheng and Wang, Yujie and Nie, Xiaonan and Zhang, Hailin and Miao, Xupeng and Cui, Bin},
  title = {{Enabling Parallelism Hot Switching for Efficient Training of Large Language Models}},
  booktitle = {Proceedings of the 30th ACM Symposium on Operating Systems Principles (SOSP)},
  year = {2024}
}

@inproceedings{vllm,
  author = {Woosuk Kwon and Zhuohan Li and Siyuan Zhuang and Ying Sheng and Lianmin Zheng and Cody Hao Yu and Joseph E. Gonzalez and Hao Zhang and Ion Stoica},
  title = {{Efficient Memory Management for Large Language Model Serving with PagedAttention}},
  booktitle = {Proceedings of the 29th ACM Symposium on Operating Systems Principles (SOSP)},
  year = {2023}
}

@inproceedings{sglang,
  author = {Lianmin Zheng and Liangsheng Yin and Zhiqiang Xie and Chuyue Sun and Jeff Huang and Cody Hao Yu and Shiyi Cao and Christos Kozyrakis and Ion Stoica and Joseph E. Gonzalez and Clark Barrett and Ying Sheng},
  title = {{SGLang: Efficient Execution of Structured Language Model Programs}},
  booktitle = {Proceedings of the Advances in Neural Information Processing Systems 37 (NeurIPS)},
  year = {2024}
}

@inproceedings{areal,
  author = {Wei Fu and Jiaxuan Gao and Xujie Shen and Chen Zhu and Zhiyu Mei and Chuyi He and Shusheng Xu and Guo Wei and Jun Mei and Jiashu Wang and Tongkai Yang and Binhang Yuan and Yi Wu},
  title = {{AReaL: A Large-Scale Asynchronous Reinforcement Learning System for Language Reasoning}},
  booktitle = {Proceedings of the Advances in Neural Information Processing Systems 38 (NeurIPS)},
  year = {2025}
}

@article{streamrl,
  author = {Yinmin Zhong and Zili Zhang and Xiaoniu Song and Hanpeng Hu and Chao Jin and Bingyang Wu and Nuo Chen and Yukun Chen and Yu Zhou and Changyi Wan and Hongyu Zhou and Yimin Jiang and Yibo Zhu and Daxin Jiang},
  title = {{StreamRL: Scalable, Heterogeneous, and Elastic RL for LLMs with Disaggregated Stream Generation}},
  journal = {arXiv preprint arXiv:2504.15930},
  year = {2025}
}

@inproceedings{MLSYS2023_5616d34c,
  author = {Hwang, Changho and Cui, Wei and Xiong, Yifan and Yang, Ziyue and Liu, Ze and Hu, Han and Wang, Zilong and Salas, Rafael and Jose, Jithin and Ram, Prabhat and Chau, HoYuen and Cheng, Peng and Yang, Fan and Yang, Mao and Xiong, Yongqiang},
  title = {{Tutel: Adaptive Mixture-of-Experts at Scale}},
  booktitle = {Proceedings of the 6th Conference on Machine Learning and Systems (MLSys)},
  year = {2023}
}

@inproceedings{10.5555/3571885.3571946,
  author = {Aminabadi, Reza Yazdani and Rajbhandari, Samyam and Awan, Ammar Ahmad and Li, Cheng and Li, Du and Zheng, Elton and Ruwase, Olatunji and Smith, Shaden and Zhang, Minjia and Rasley, Jeff and He, Yuxiong},
  title = {{DeepSpeed-Inference: Enabling Efficient Inference of Transformer Models at Unprecedented Scale}},
  booktitle = {Proceedings of the International Conference for High Performance Computing, Networking, Storage and Analysis (SC)},
  year = {2022}
}

@inproceedings{Singh_2023,
  author = {Singh, Siddharth and Ruwase, Olatunji and Awan, Ammar Ahmad and Rajbhandari, Samyam and He, Yuxiong and Bhatele, Abhinav},
  title = {{A Hybrid Tensor-Expert-Data Parallelism Approach to Optimize Mixture-of-Experts Training}},
  booktitle = {Proceedings of the 37th ACM International Conference on Supercomputing (ICS)},
  year = {2023}
}

@inproceedings{spotserve,
  author = {Miao, Xupeng and Shi, Chunan and Duan, Jiangfei and Xi, Xiaoli and Lin, Dahua and Cui, Bin and Jia, Zhihao},
  title = {{SpotServe: Serving Generative Large Language Models on Preemptible Instances}},
  booktitle = {Proceedings of the 29th International Conference on Architectural Support for Programming Languages and Operating Systems (ASPLOS)},
  year = {2024}
}

@inproceedings{lina,
  author = {Jiamin Li and Yimin Jiang and Yibo Zhu and Cong Wang and Hong Xu},
  title = {{Accelerating Distributed MoE Training and Inference with Lina}},
  booktitle = {Proceedings of the 2023 USENIX Annual Technical Conference (USENIX ATC)},
  year = {2023}
}

@inproceedings{orca,
  author = {Gyeong-In Yu and Joo Seong Jeong and Geon-Woo Kim and Soojeong Kim and Byung-Gon Chun},
  title = {{Orca: A Distributed Serving System for Transformer-Based Generative Models}},
  booktitle = {Proceedings of the 16th USENIX Symposium on Operating Systems Design and Implementation (OSDI)},
  year = {2022}
}

@inproceedings{flashattention,
  author = {Dao, Tri and Fu, Daniel Y. and Ermon, Stefano and Rudra, Atri and R{\'e}, Christopher},
  title = {{FlashAttention: Fast and Memory-Efficient Exact Attention with IO-Awareness}},
  booktitle = {Proceedings of the Advances in Neural Information Processing Systems 35 (NeurIPS)},
  year = {2022}
}

@inproceedings{flashinfer,
  author = {Ye, Zihao and Chen, Lequn and Lai, Ruihang and Lin, Wuwei and Zhang, Yineng and Wang, Stephanie and Chen, Tianqi and Kasikci, Baris and Grover, Vinod and Krishnamurthy, Arvind and Ceze, Luis},
  title = {{FlashInfer: Efficient and Customizable Attention Engine for LLM Inference Serving}},
  booktitle = {Proceedings of the 8th Conference on Machine Learning and Systems (MLSys)},
  year = {2025}
}

@article{ucclep,
  author = {Ziming Mao and Yihan Zhang and Chihan Cui and Zhen Huang and Kaichao You and Zhongjie Chen and Zhiying Xu and Zhenyu Gu and Scott Shenker and Costin Raiciu and Yang Zhou and Ion Stoica},
  title = {{UCCL-EP: Portable Expert-Parallel Communication}},
  journal = {arXiv preprint arXiv:2512.19849},
  year = {2026}
}

@article{lin2025haphybridadaptiveparallelism,
  author = {Haoran Lin and Xianzhi Yu and Kang Zhao and Han Bao and Zongyuan Zhan and Ting Hu and Wulong Liu and Zekun Yin and Xin Li and Weiguo Liu},
  title = {{HAP: Hybrid Adaptive Parallelism for Efficient Mixture-of-Experts Inference}},
  journal = {arXiv preprint arXiv:2508.19373},
  year = {2025}
}

@inproceedings{rollpacker,
  author = {Wei Gao and Yuheng Zhao and Dakai An and Tianyuan Wu and Lunxi Cao and Shaopan Xiong and Ju Huang and Weixun Wang and Siran Yang and Wenbo Su and Jiamang Wang and Lin Qu and Bo Zheng and Wei Wang},
  title = {{RollPacker: Taming Long-Tail Rollouts for RL Post-Training with Tail Batching}},
  booktitle = {Proceedings of the 23rd USENIX Symposium on Networked Systems Design and Implementation (NSDI)},
  year = {2026}
}

@inproceedings{TLT,
  author = {Qinghao Hu and Shang Yang and Junxian Guo and Xiaozhe Yao and Yujun Lin and Yuxian Gu and Han Cai and Chuang Gan and Ana Klimovic and Song Han},
  title = {{Taming the Long-Tail: Efficient Reasoning RL Training with Adaptive Drafter}},
  booktitle = {Proceedings of the 31st ACM International Conference on Architectural Support for Programming Languages and Operating Systems (ASPLOS)},
  year = {2026}
}

@article{XiaomiR3,
  author = {Wenhan Ma and Hailin Zhang and Liang Zhao and Yifan Song and Yudong Wang and Zhifang Sui and Fuli Luo},
  title = {{Stabilizing MoE Reinforcement Learning by Aligning Training and Inference Routers}},
  journal = {arXiv preprint arXiv:2510.11370},
  year = {2025}
}

@article{longcatdora,
  author = {Tianhao Hu and Xiangcheng Liu and Youshao Xiao and Yang Zheng and Xuan Huang and Jinrui Ding and Yufei Zhang and Tao Liang and Hongyu Zang and Quan Chen and Yueqing Sun and Wenjie Shi and Chao Zhang and Wei Wang and Qi Gu and Yerui Sun and Yucheng Xie and Xunliang Cai},
  title = {{DORA: A Scalable Asynchronous Reinforcement Learning System for Language Model Training}},
  journal = {arXiv preprint arXiv:2604.26256},
  year = {2026}
}

@inproceedings{laminar,
  author = {Sheng, Guangming and Tong, Yuxuan and Wan, Borui and Zhang, Wang and Jia, Chaobo and Wu, Xibin and Wu, Yuqi and Li, Xiang and Zhang, Chi and Peng, Yanghua and Lin, Haibin and Liu, Xin and Wu, Chuan},
  title = {{Laminar: A Scalable Asynchronous RL Post-Training Framework}},
  booktitle = {Proceedings of the 21st European Conference on Computer Systems (EuroSys)},
  year = {2026}
}

@inproceedings{deepmath,
  author = {Zhiwei He and Tian Liang and Jiahao Xu and Qiuzhi Liu and Xingyu Chen and Yue Wang and Linfeng Song and Dian Yu and Zhenwen Liang and Wenxuan Wang and Zhuosheng Zhang and Rui Wang and Zhaopeng Tu and Haitao Mi and Dong Yu},
  title = {{DeepMath-103K: A Large-Scale, Challenging, Decontaminated, and Verifiable Mathematical Dataset for Advancing Reasoning}},
  booktitle = {Proceedings of the 14th International Conference on Learning Representations (ICLR)},
  year = {2026}
}

@article{grpo,
  author = {Zhihong Shao and Peiyi Wang and Qihao Zhu and Runxin Xu and Junxiao Song and Xiao Bi and Haowei Zhang and Mingchuan Zhang and Y. K. Li and Y. Wu and Daya Guo},
  title = {{DeepSeekMath: Pushing the Limits of Mathematical Reasoning in Open Language Models}},
  journal = {arXiv preprint arXiv:2402.03300},
  year = {2024}
}

@article{qwen3,
  author = {An Yang and Anfeng Li and Baosong Yang and Beichen Zhang and Binyuan Hui and Bo Zheng and Bowen Yu and Chang Gao and Chengen Huang and Chenxu Lv and Chujie Zheng and Dayiheng Liu and Fan Zhou and Fei Huang and Feng Hu and Hao Ge and Haoran Wei and Huan Lin and Jialong Tang and Jian Yang and Jianhong Tu and Jianwei Zhang and Jianxin Yang and Jiaxi Yang and Jing Zhou and Jingren Zhou and Junyang Lin and Kai Dang and Keqin Bao and Kexin Yang and Le Yu and Lianghao Deng and Mei Li and Mingfeng Xue and Mingze Li and Pei Zhang and Peng Wang and Qin Zhu and Rui Men and Ruize Gao and Shixuan Liu and Shuang Luo and Tianhao Li and Tianyi Tang and Wenbiao Yin and Xingzhang Ren and Xinyu Wang and Xinyu Zhang and Xuancheng Ren and Yang Fan and Yang Su and Yichang Zhang and Yinger Zhang and Yu Wan and Yuqiong Liu and Zekun Wang and Zeyu Cui and Zhenru Zhang and Zhipeng Zhou and Zihan Qiu},
  title = {{Qwen3 Technical Report}},
  journal = {arXiv preprint arXiv:2505.09388},
  year = {2025}
}

@article{deepseekv3,
  author = {DeepSeek-AI and Aixin Liu and Bei Feng and Bing Xue and Bingxuan Wang and Bochao Wu and Chengda Lu and Chenggang Zhao and Chengqi Deng and Chenyu Zhang and Chong Ruan and Damai Dai and Daya Guo and Dejian Yang and Deli Chen and Dongjie Ji and Erhang Li and Fangyun Lin and Fucong Dai and Fuli Luo and Guangbo Hao and Guanting Chen and Guowei Li and H. Zhang and Han Bao and Hanwei Xu and Haocheng Wang and Haowei Zhang and Honghui Ding and Huajian Xin and Huazuo Gao and Hui Li and Hui Qu and J. L. Cai and Jian Liang and Jianzhong Guo and Jiaqi Ni and Jiashi Li and Jiawei Wang and Jin Chen and Jingchang Chen and Jingyang Yuan and Junjie Qiu and Junlong Li and Junxiao Song and Kai Dong and Kai Hu and Kaige Gao and Kang Guan and Kexin Huang and Kuai Yu and Lean Wang and Lecong Zhang and Lei Xu and Leyi Xia and Liang Zhao and Litong Wang and Liyue Zhang and Meng Li and Miaojun Wang and Mingchuan Zhang and Minghua Zhang and Minghui Tang and Mingming Li and Ning Tian and Panpan Huang and Peiyi Wang and Peng Zhang and Qiancheng Wang and Qihao Zhu and Qinyu Chen and Qiushi Du and R. J. Chen and R. L. Jin and Ruiqi Ge and Ruisong Zhang and Ruizhe Pan and Runji Wang and Runxin Xu and Ruoyu Zhang and Ruyi Chen and S. S. Li and Shanghao Lu and Shangyan Zhou and Shanhuang Chen and Shaoqing Wu and Shengfeng Ye and Shengfeng Ye and Shirong Ma and Shiyu Wang and Shuang Zhou and Shuiping Yu and Shunfeng Zhou and Shuting Pan and T. Wang and Tao Yun and Tian Pei and Tianyu Sun and W. L. Xiao and Wangding Zeng and Wanjia Zhao and Wei An and Wen Liu and Wenfeng Liang and Wenjun Gao and Wenqin Yu and Wentao Zhang and X. Q. Li and Xiangyue Jin and Xianzu Wang and Xiao Bi and Xiaodong Liu and Xiaohan Wang and Xiaojin Shen and Xiaokang Chen and Xiaokang Zhang and Xiaosha Chen and Xiaotao Nie and Xiaowen Sun and Xiaoxiang Wang and Xin Cheng and Xin Liu and Xin Xie and Xingchao Liu and Xingkai Yu and Xinnan Song and Xinxia Shan and Xinyi Zhou and Xinyu Yang and Xinyuan Li and Xuecheng Su and Xuheng Lin and Y. K. Li and Y. Q. Wang and Y. X. Wei and Y. X. Zhu and Yang Zhang and Yanhong Xu and Yanhong Xu and Yanping Huang and Yao Li and Yao Zhao and Yaofeng Sun and Yaohui Li and Yaohui Wang and Yi Yu and Yi Zheng and Yichao Zhang and Yifan Shi and Yiliang Xiong and Ying He and Ying Tang and Yishi Piao and Yisong Wang and Yixuan Tan and Yiyang Ma and Yiyuan Liu and Yongqiang Guo and Yu Wu and Yuan Ou and Yuchen Zhu and Yuduan Wang and Yue Gong and Yuheng Zou and Yujia He and Yukun Zha and Yunfan Xiong and Yunxian Ma and Yuting Yan and Yuxiang Luo and Yuxiang You and Yuxuan Liu and Yuyang Zhou and Z. F. Wu and Z. Z. Ren and Zehui Ren and Zhangli Sha and Zhe Fu and Zhean Xu and Zhen Huang and Zhen Zhang and Zhenda Xie and Zhengyan Zhang and Zhewen Hao and Zhibin Gou and Zhicheng Ma and Zhigang Yan and Zhihong Shao and Zhipeng Xu and Zhiyu Wu and Zhongyu Zhang and Zhuoshu Li and Zihui Gu and Zijia Zhu and Zijun Liu and Zilin Li and Ziwei Xie and Ziyang Song and Ziyi Gao and Zizheng Pan},
  title = {{DeepSeek-V3 Technical Report}},
  journal = {arXiv preprint arXiv:2412.19437},
  year = {2025}
}

@article{gptoss,
  author = {{OpenAI} and Sandhini Agarwal and Lama Ahmad and Jason Ai and Sam Altman and Andy Applebaum and Edwin Arbus and Rahul K. Arora and Yu Bai and Bowen Baker and Haiming Bao and Boaz Barak and Ally Bennett and Tyler Bertao and Nivedita Brett and Eugene Brevdo and Greg Brockman and Sebastien Bubeck and Che Chang and Kai Chen and Mark Chen and Enoch Cheung and Aidan Clark and Dan Cook and Marat Dukhan and Casey Dvorak and Kevin Fives and Vlad Fomenko and Timur Garipov and Kristian Georgiev and Mia Glaese and Tarun Gogineni and Adam Goucher and Lukas Gross and Katia Gil Guzman and John Hallman and Jackie Hehir and Johannes Heidecke and Alec Helyar and Haitang Hu and Romain Huet and Jacob Huh and Saachi Jain and Zach Johnson and Chris Koch and Irina Kofman and Dominik Kundel and Jason Kwon and Volodymyr Kyrylov and Elaine Ya Le and Guillaume Leclerc and James Park Lennon and Scott Lessans and Mario Lezcano-Casado and Yuanzhi Li and Zhuohan Li and Ji Lin and Jordan Liss and Lily Liu and Jiancheng Liu and Kevin Lu and Chris Lu and Zoran Martinovic and Lindsay McCallum and Josh McGrath and Scott McKinney and Aidan McLaughlin and Song Mei and Steve Mostovoy and Tong Mu and Gideon Myles and Alexander Neitz and Alex Nichol and Jakub Pachocki and Alex Paino and Dana Palmie and Ashley Pantuliano and Giambattista Parascandolo and Jongsoo Park and Leher Pathak and Carolina Paz and Ludovic Peran and Dmitry Pimenov and Michelle Pokrass and Elizabeth Proehl and Huida Qiu and Gaby Raila and Filippo Raso and Hongyu Ren and Kimmy Richardson and David Robinson and Bob Rotsted and Hadi Salman and Suvansh Sanjeev and Max Schwarzer and D. Sculley and Harshit Sikchi and Kendal Simon and Karan Singhal and Yang Song and Dane Stuckey and Zhiqing Sun and Philippe Tillet and Sam Toizer and Foivos Tsimpourlas and Nikhil Vyas and Eric Wallace and Xin Wang and Miles Wang and Olivia Watkins and Kevin Weil and Amy Wendling and Kevin Whinnery and Cedric Whitney and Hannah Wong and Lin Yang and Yu Yang and Michihiro Yasunaga and Kristen Ying and Wojciech Zaremba and Wenting Zhan and Cyril Zhang and Brian Zhang and Eddie Zhang and Shengjia Zhao},
  title = {{gpt-oss-120b \& gpt-oss-20b Model Card}},
  journal = {arXiv preprint arXiv:2508.10925},
  year = {2025}
}

@inproceedings{splitwise,
  author = {Patel, Pratyush and Choukse, Esha and Zhang, Chaojie and Shah, Aashaka and Goiri, \'{I}\~{n}igo and Maleki, Saeed and Bianchini, Ricardo},
  title = {{Splitwise: Efficient Generative LLM Inference Using Phase Splitting}},
  booktitle = {Proceedings of the 51st International Symposium on Computer Architecture (ISCA)},
  year = {2024}
}

@article{kimik1.5,
  author = {Kimi Team and Angang Du and Bofei Gao and Bowei Xing and Changjiu Jiang and Cheng Chen and Cheng Li and Chenjun Xiao and Chenzhuang Du and Chonghua Liao and Chuning Tang and Congcong Wang and Dehao Zhang and Enming Yuan and Enzhe Lu and Fengxiang Tang and Flood Sung and Guangda Wei and Guokun Lai and Haiqing Guo and Han Zhu and Hao Ding and Hao Hu and Hao Yang and Hao Zhang and Haotian Yao and Haotian Zhao and Haoyu Lu and Haoze Li and Haozhen Yu and Hongcheng Gao and Huabin Zheng and Huan Yuan and Jia Chen and Jianhang Guo and Jianlin Su and Jianzhou Wang and Jie Zhao and Jin Zhang and Jingyuan Liu and Junjie Yan and Junyan Wu and Lidong Shi and Ling Ye and Longhui Yu and Mengnan Dong and Neo Zhang and Ningchen Ma and Qiwei Pan and Qucheng Gong and Shaowei Liu and Shengling Ma and Shupeng Wei and Sihan Cao and Siying Huang and Tao Jiang and Weihao Gao and Weimin Xiong and Weiran He and Weixiao Huang and Weixin Xu and Wenhao Wu and Wenyang He and Xianghui Wei and Xianqing Jia and Xingzhe Wu and Xinran Xu and Xinxing Zu and Xinyu Zhou and Xuehai Pan and Y. Charles and Yang Li and Yangyang Hu and Yangyang Liu and Yanru Chen and Yejie Wang and Yibo Liu and Yidao Qin and Yifeng Liu and Ying Yang and Yiping Bao and Yulun Du and Yuxin Wu and Yuzhi Wang and Zaida Zhou and Zhaoji Wang and Zhaowei Li and Zhen Zhu and Zheng Zhang and Zhexu Wang and Zhilin Yang and Zhiqi Huang and Zihao Huang and Ziyao Xu and Zonghan Yang and Zongyu Lin},
  title = {{Kimi k1.5: Scaling Reinforcement Learning with LLMs}},
  journal = {arXiv preprint arXiv:2501.12599},
  year = {2025}
}

@inproceedings{rlhfuse,
  author = {Zhong, Yinmin and Zhang, Zili and Wu, Bingyang and Liu, Shengyu and Chen, Yukun and Wan, Changyi and Hu, Hanpeng and Xia, Lei and Ming, Ranchen and Zhu, Yibo and Jin, Xin},
  title = {{Optimizing RLHF Training for Large Language Models with Stage Fusion}},
  booktitle = {Proceedings of the 22nd USENIX Symposium on Networked Systems Design and Implementation (NSDI)},
  year = {2025}
}

@misc{azurellmtrace,
  author = {{Microsoft Azure}},
  title = {{Azure LLM Inference Trace}},
  year = {2024},
  note = {\url{https://github.com/Azure/AzurePublicDataset/blob/master/AzureLLMInferenceDataset2024.md}}
}

@misc{semianalysis_inferencex,
  author = {{SemiAnalysis}},
  title = {{InferenceX: LLM Inference Performance Benchmarks}},
  year = {2025},
  note = {\url{https://inferencex.semianalysis.com/inference}}
}

@misc{tensorrtllm,
  author = {{NVIDIA}},
  title = {{TensorRT-LLM: A TensorRT Toolbox for Optimized Large Language Model Inference}},
  year = {2023},
  note = {\url{https://github.com/NVIDIA/TensorRT-LLM}}
}

\clearpage

\appendix
\section{Rollout Workload Distribution}
\label{app:rollout-dist}

\begin{figure}[h]
	\centering
	\includegraphics[width=\columnwidth]{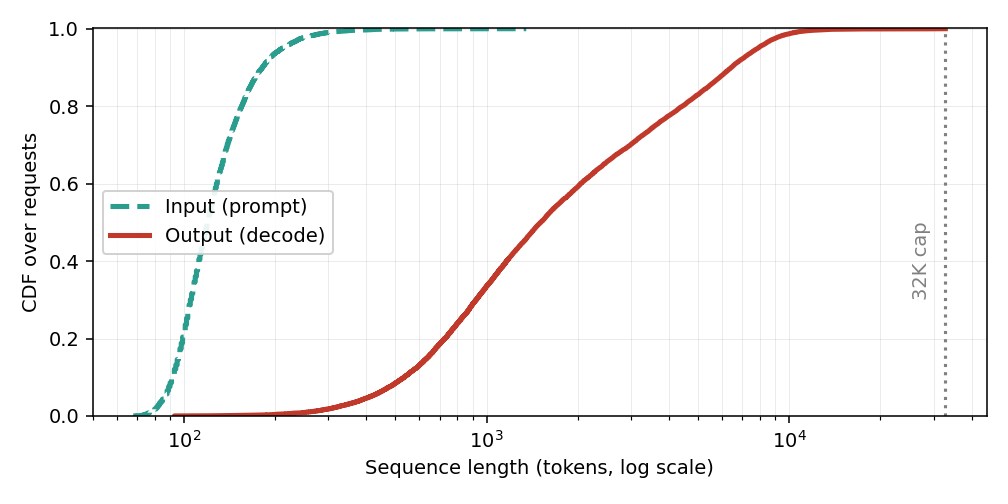}
	\caption{Input (dashed) and output (solid) token-length CDFs over the nine DeepMath rollout steps (18{,}432 requests). Grey dotted line marks the 32k decode cap.}
	\label{fig:eval-rollout-dist}
\end{figure}

\figref{fig:eval-rollout-dist} plots the input and output token-length CDFs pooled over the nine DeepMath rollout steps (18{,}432 requests).
The two are sharply asymmetric: inputs are short and tightly clustered, while outputs are long and heavy-tailed, running out to the 32k decode cap.
This output tail is what makes a rollout step's active batch decay slowly, the property the rollout evaluation (\secref{sec:eval-rollout}) exploits.

\section{End-to-End Training Projection}
\label{app:e2e-projection}

Our per-step rollout speedup carries to end-to-end training time differently depending on how the RL framework overlaps generation with policy updates (\secref{sec:bg-workloads}).
Synchronous on-policy training waits for the full rollout step before each update~\cite{hybridflow,hu2024openrlhf,rlhfuse}, so by Amdahl's law a per-step speedup \(s\) over a rollout fraction \(f\) gives \(1/((1{-}f)+f/s)\) end-to-end, and reported fractions of 65--85\%~\cite{rollpacker,TLT} turn our 1.16--1.25\(\times\) into 1.10--1.20\(\times\).
One-step-stale overlap is rollout-bound, so the gain passes through nearly one-to-one~\cite{streamrl,areal}, while partial-rollout~\cite{kimik1.5,areal} and trajectory-level asynchronous~\cite{laminar} schemes reshape the per-step workload and need a real integrated run to quantify.
We report the per-step speedup as our measured result and treat these end-to-end figures as projections.

\section{Runtime Memory Breakdown}
\label{app:memory}

\tabref{tab:eval-memory} details the per-GPU runtime state of \figref{fig:eval-memory}, the footprint outside weights, KV cache, and the dual-mode buffer.
The 2.8\,GB dual-mode buffer itself divides into 1.7\,GB of TP-mode \texttt{qkv\_proj} and \texttt{o\_proj} shards held alongside the full EP copies and 1.1\,GB for one spare physical layer that stages the per-layer transfer.
\textsc{TP}'s larger activation workspaces and CUDA graphs follow from sizing for its full 2{,}048-request batch, where \sysname and \textsc{EP} size for a per-rank batch of 256.
The 12.7\,GB weight gap between data-parallel attention (\sysname, \textsc{EP}) and sharded attention (\textsc{TP}) comes from the attention projections.
Qwen3-235B's attention uses 64 heads of dimension 128, so its q/k/v/o projections total roughly 13\,GB in BF16 across its 94 layers.
\textsc{TP} holds one-eighth per GPU, \textsc{EP} a full copy, with the per-rank embedding and LM head making up the small remainder.
\textsc{TP}'s smaller weight footprint leaves it the most room for KV cache, but the extra capacity buys little throughput, since \textsc{TP}'s decode is bound by compute and communication rather than concurrency.

\begin{table}[h]
	\centering
	\caption{Runtime state per GPU (GB), the footprint outside weights, KV cache, and the dual-mode buffer (static allocation post-capture). Columns may not sum to the total due to rounding.}
	\label{tab:eval-memory}
	\begin{tabular}{lccc}
		\toprule
		Component & \textsc{TP} & \textsc{EP} & \sysname \\
		\midrule
		Activation workspaces          & 7.6  & 3.1 & 3.0 \\
		CUDA graphs                    & 1.8  & 0.8 & 0.9 \\
		EP infra (DeepEP\,+\,NVSHMEM)  & ---  & 2.2 & 2.2 \\
		NCCL communicators             & 0.6  & 0.6 & 0.6 \\
		Other (NCCL VMM, cuBLAS, IPC)  & 2.9  & 1.4 & 1.7 \\
		\midrule
		Total runtime                  & 12.7 & 8.1 & 8.3 \\
		\bottomrule
	\end{tabular}
\end{table}



\end{document}